%% file: main-fullpage.tex
\documentclass[11pt]{article}
\pdfoutput=1
\usepackage{hyperref}
    
\usepackage[margin=1in]{geometry}
\usepackage[utf8]{inputenc}
\usepackage[pdftex]{graphicx}
\usepackage{microtype}
\usepackage[all=normal,paragraphs=tight,bibliography=tight,floats=tight]{savetrees}
\usepackage[english]{babel}
\usepackage[cmex10]{amsmath}
\usepackage{todonotes}
\usepackage{amsthm}
\usepackage{amssymb}
\usepackage{amsfonts}
\usepackage{xspace}
\usepackage{cite}
\usepackage{comment}
\usepackage{letltxmacro}
\usepackage{enumerate}
\usepackage{url}
\usepackage{textcomp}
\usepackage{tikz}
\usepackage{boxedminipage}
\usepackage{marvosym}

\usepackage{verbatim}
\usepackage{graphics}
\usepackage{listings}
\usepackage{subfig}
\usepackage{comment}

\newtheorem{theorem}{Theorem}
\newtheorem{lemma}[theorem]{Lemma}
\newtheorem{corollary}[theorem]{Corollary}

\theoremstyle{definition}
\newtheorem{definition}[theorem]{Definition}

\newtheorem{claim}[theorem]{Claim}
\newtheorem{hypothesis}[theorem]{Hypothesis}

\def\cqedsymbol{\ifmmode$\lrcorner$\else{\unskip\nobreak\hfil
\penalty50\hskip1em\null\nobreak\hfil$\lrcorner$
\parfillskip=0pt\finalhyphendemerits=0\endgraf}\fi} 

\newcommand{\cqed}{\renewcommand{\qed}{\cqedsymbol}}

\newcommand{\executeiffilenewer}[3]{%
\ifnum\pdfstrcmp{\pdffilemoddate{#1}}%
{\pdffilemoddate{#2}}>0%
{\immediate\write18{#3}}\fi%
} 
\usetikzlibrary{arrows,shapes}
\usetikzlibrary{trees}

\newcommand{\minhorn}{{\sc{Min Horn Deletion}}\xspace}
\newcommand{\Hfreedel}{{\sc$H$-free Edge Deletion}\xspace}

\newcommand{\gHfreedel}{{\sc Sandwich $H$-free Edge Deletion}\xspace}
\newcommand{\gHfreecom}{{\sc Sandwich $H$-free Edge Completion}\xspace}
\newcommand{\Hfreecom}{{\sc $H$-free Edge Completion}\xspace}

\newcommand{\Hfreedelcom}{{\sc $H$-free Edge Deletion (Completion)}\xspace}

\newcommand{\minones}{{\sc MinOnes($\mathcal{F}$)}\xspace}
\newcommand{\minoness}{{\sc MinOnes($\mathcal{F'}$)}\xspace}
\newcommand{\minonesn}{{\sc MinOnes($\mathcal{F}_n''$)}\xspace}
\newcommand{\quarkndel}{{\sc Quarantined $K_n\setminus e$-free Edge Deletion}\xspace}
\newcommand{\kndel}{{\sc $K_n\setminus e$-free Edge Deletion}\xspace}

\newcommand{\quar}{{\sc Quarantined $H$-free Edge Deletion}\xspace}

\newcommand{\nn}[1]{\neg{#1} }	
\newcommand{\tn}[1]{\footnotesize{#1}}
\newcommand{\tnn}[1]{\tiny{#1}}
\newcommand{\poly}{\mathrm{poly}}
\newcommand{\OPT}{\mathsf{OPT}}
\def\threesat{\text{\sc 3SAT}\xspace}

\newcommand{\Oh}{{\mathcal{O}}}

\title{Hardness of approximation for $H$-free edge modification problems\thanks{The research of Mi. Pilipczuk is supported by Polish National Science Centre grant UMO-2013/11/D/ST6/03073.
Mi. Pilipczuk is also supported by the Foundation for Polish Science (FNP) via the START stipend programme.}}
\author{
  Ivan Bliznets\thanks{
    St.~Petersburg Department of Steklov Institute of Mathematics, Russia, \texttt{iabliznets@gmail.com}.
  }
  \and 
  Marek Cygan\thanks{
    Institute of Informatics, University of Warsaw, Poland, \texttt{cygan@mimuw.edu.pl}.
  }
  \and 
  Pawe\l{} Komosa\thanks{
    Institute of Informatics, University of Warsaw, Poland, \texttt{p.komosa@mimuw.edu.pl}.
  }
  \and 
  Micha\l{} Pilipczuk\thanks{
    Institute of Informatics, University of Warsaw, Poland, \texttt{michal.pilipczuk@mimuw.edu.pl}.
  }
}

\date{}

\begin{document}

\begin{titlepage}
\def\thepage{}
\thispagestyle{empty}
\maketitle

\input{abstract}
\end{titlepage}

  \input intro
  \input preliminaries
  \input reduction

\input specific
  \input conclusion

 \bibliographystyle{abbrv}

  \bibliography{main-fullpage}

  \appendix
  \section{Omitted proofs}

  In the following we present the proof of Theorem~\ref{thm:main} for \Hfreecom.

  \input completion

\end{document}

%% file: abstract.tex
\begin{abstract}
The \Hfreedel problem asks, for a given graph $G$ and integer $k$, 
whether it is possible to delete at most $k$ edges from $G$ to make it {\em{$H$-free}}, that is, not containing $H$ as an induced subgraph.
The \Hfreecom problem is defined similarly, but we add edges instead of deleting them.
The study of these two problem families has recently been the subject of intensive studies from the point of view of parameterized complexity and kernelization.
In particular, it was shown that the problems do not admit polynomial kernels (under plausible complexity assumptions) for almost all graphs $H$, 
with several important exceptions occurring when the class of $H$-free graphs exhibits some structural properties.

In this work we complement the parameterized study of edge modification problems to $H$-free graphs by considering their approximability.
We prove that whenever $H$ is $3$-connected and has at least two non-edges, then both \Hfreedel and \Hfreecom are very hard to approximate: 
they do not admit $\poly(\OPT)$-approximation in polynomial time, unless $\mathrm{P}=\mathrm{NP}$, or even in time subexponential in $\OPT$, unless the Exponential Time Hypothesis fails.
The assumption of the existence of two non-edges appears to be important: we show that whenever $H$ is a complete graph without one edge, 
then \Hfreedel is tightly connected to the \minhorn problem, whose approximability is still open.
Finally, in an attempt to extend our hardness results beyond $3$-connected graphs, we consider the cases of $H$ being a path or a cycle, and we achieve an almost complete dichotomy there.
\end{abstract}

%% file: intro.tex
\section{Introduction}

We consider the following general setting of {\em{graph modification problems}}: given a graph $G$, one would like to modify $G$ as little as possible in order to make it satisfy some fixed property of
global nature. 
Motivated by applications in de-noising data derived from imprecise experimental measurements, graph modification problems occupy a prominent role in the field of parameterized complexity and kernelization.
This is because the allowed number of modifications usually can be assumed to be small compared to the total instance size, which exactly fits the motivation of considering it as the parameter of the instance.

Moving to the formal setting, consider some hereditary class of graphs $\Pi$, that is, a class closed under taking induced subgraphs. 
For such a class $\Pi$, we can define several problems depending on the set of allowed modifications.
In each case the input consists of a graph $G$ and integer $k$, and the question is whether one can apply at most $k$ modification to $G$ so that it falls into class $\Pi$.
In this paper we will consider {\em{deletion}} and {\em{completion}} problems, where we are allowed only to delete edges, respectively only to add edges.
However, other studied variants include {\em{vertex deletion}} problems (the allowed modification is removal of a vertex) and {\em{editing}} problems (both edge deletions and completions are allowed).
Moreover, we restrict ourselves to classes $\Pi$ characterized by one forbidden induced subgraph $H$.
In other words, $\Pi$ is the class of $H$-free graphs, that is, graphs that do not contain $H$ as an induced subgraph ($H$ is assumed to be constant).

The study of the parameterized complexity of \Hfreedel and \Hfreecom focused on two aspects: designing fixed-parameter algorithms and kernelization procedures.
The classic observation of Cai~\cite{cai1996fixed} shows that \Hfreedelcom can be both solved in time $c^k\cdot n^{\Oh(1)}$ for some constant $c$ depending only on $H$, using a straightforward branching strategy.
However, for several completion problems related to chordal graphs and their subclasses, like (proper) interval graphs or trivially perfect graphs, 
one can design {\em{subexponential parameterized algorithms}}, typically with the running time of $2^{\Oh(\sqrt{k}\log k)}\cdot n^{\Oh(1)}$.
The study of this surprising subexponential phenomenon, and of its limits, has recently been the subject of intensive studies; we refer to the introductory section of~\cite{BliznetsCKMP16} for more details.
However, for the vast majority of graphs $H$, the running time of the form $c^k\cdot n^{\Oh(1)}$ is essentially the best one can hope for \Hfreedelcom.
Indeed, Aravind et al.~\cite{AravindSS15} proved that, whenever $H$ has at least two edges, then \Hfreedel is NP-hard and has no $2^{o(k)}\cdot n^{\Oh(1)}$-time algorithm unless the Exponential Time Hypothesis fails,
and the same result holds for \Hfreecom whenever $H$ has at least two non-edges. The remaining cases are easily seen to be polynomial-time solvable, so this establishes a full dichotomy.

Another interesting aspect of graph modification problems is their kernelization complexity. Recall that a {\em{polynomial kernel}} for a parameterized problem is a polynomial-time algorithm that,
given an instance of the problem with parameter $k$, reduces it to another instance of the same problem that has size bounded polynomially in $k$. 
While every {\sc{$H$-Free Vertex Deletion}} problem admits a simple polynomial kernel by a reduction to the {\sc{$d$-Hitting Set}} problem (for $d=|V(H)|$), the situation for edge deletion and edge completion
problems is much more complex. This is because the removal/addition of some edge may create new induced copies of $H$ that were originally not present, 
and hence the obstacles can ``propagate'' in the graph. 
In fact, a line of work~\cite{CaiC15,cai2012master,GuillemotHPP13,KratschW13} showed that, unless $\mathrm{NP}\subseteq \mathrm{coNP}/\poly$,
polynomial kernels for the \Hfreedelcom problems exist only for very simple graphs $H$, for which the class of $H$-free graphs exhibits some structural property. 
This line culminated in the work of Cai and Cai~\cite{CaiC15,cai2012master}, who attempted to obtain a complete dichotomy.
While this goal was not fully achieved and there are some cases missing, the obtained complexity picture explains the general situation very well. 
For example, Cai and Cai~\cite{CaiC15,cai2012master} showed that polynomial kernels do not exist (under $\mathrm{NP}\nsubseteq \mathrm{coNP}/\poly$) for the \Hfreedelcom problems whenever $H$ is $3$-connected and has at least $2$ non-edges.
Nontrivial positive cases include e.g. $H$ being a path on $4$ vertices~\cite{GuillemotHPP13} (that is, {\sc{Cograph Edge Deletion (Completion)}}), and $H$ being a $K_4$ minus one edge~\cite{cai2012master}
(that is, {\sc{Diamond-free Edge Deletion}}).
One of the most prominent open cases left is the kernelization complexity of {\sc{Claw-Free Edge Deletion}}~\cite{CaiC15,CyganPPLW15}.

\paragraph*{Our motivation and results.} The starting point of our work is the realization that the propagational character of \Hfreedelcom, which is the basic explanation of its apparent kernelization hardness,
also makes the greedy approach to approximation incorrect. One cannot greedily remove all the edges of any copy of $H$ in the graph, because removing an edge does not necessarily always help: it
may create new copies of $H$ in the instance. Hence, the approximation complexity of \Hfreedelcom is actually also highly unclear. 
On the other hand, the links between approximation and kernelization are well-known in parameterized complexity: 
it is often the case that a polynomial kernel for a problem can be turned into a~$\poly(\OPT)$-approximation algorithm 
(i.e. an algorithm that returns a solution of cost bounded by some polynomial function of the optimum), by just taking greedily the kernel and reverting the reduction rules.
While this intuitive link is far from being formal, and actually there are examples of problems behaving differently~\cite{GiannopoulouLSS14}, 
it is definitely the case that the combinatorial insight given by kernelization algorithms may be very useful in the approximation setting.

Therefore, we propose to study the approximability of \Hfreedelcom as well, alongside with the best possible running times of fixed-parameter algorithms and the existence of polynomial kernels.
This work is the first step in this direction.

We prove that the \Hfreedelcom problems are very hard to approximate for a vast majority of graphs $H$, which mirrors the kernelization hardness results of Cai and Cai~\cite{CaiC15,cai2012master}. 
The following theorem explains our main result formally.

\begin{theorem}\label{thm:main}
Let $H$ be a $3$-connected graph with at least two non-edges. Then, unless $\mathrm{P}=\mathrm{NP}$, neither \Hfreedel nor \Hfreecom admits a $\poly(\OPT)$-approximation algorithm running in polynomial time.
Moreover, unless the Exponential Time Hypothesis fails, neither of these problems admits even a $\poly(\OPT)$-approximation algorithm running in time $2^{o(\OPT)}\cdot n^{\Oh(1)}$.
\end{theorem}

Theorem~\ref{thm:main} makes two structural assumptions about graph $H$: that it is $3$-connected, and has at least two non-edges.
The first one is a crucial technical ingredient in the reductions, because it enables us to argue that for any vertex cut of size $2$, every copy of $H$ in the graph is completely contained on one side of the cut.
Relaxing this assumption is a major issue addressed by Cai and Cai~\cite{CaiC15,cai2012master} in their work. 
In an attempt to lift this assumption in our setting as well, we try to resolve the case of $H$ being a path or a cycle first;
this reflects the development of the story of kernelization hardness for the considered problems~\cite{cai2012master,CaiC15,GuillemotHPP13,KratschW13}.
The following theorem summarizes our results in this direction.

\begin{theorem}\label{thm:main-paths-cycles}
Let $H$ be a cycle on at least $4$ vertices or a path on at least $5$ vertices. 
Then, unless $\mathrm{P}=\mathrm{NP}$, neither \Hfreedel nor \Hfreecom admits a $\poly(\OPT)$-approximation algorithm running in polynomial time.
Moreover, unless the Exponential Time Hypothesis fails, neither of these problems admits even a $\poly(\OPT)$-approximation algorithm running in time $2^{o(\OPT)}\cdot n^{\Oh(1)}$.
\end{theorem}

Together with some easy cases and known positive results~\cite{natanzonmaster}, this gives an almost complete dichotomy for paths and cycles. 
The only missing case is {\sc{Cograph Edge Deletion}} (for $H=P_4$), for which we expect a positive answer due to the existence of a polynomial kernel~\cite{GuillemotHPP13}.
However, our preliminary attempt at lifting the kernel of Guillemot et al.~\cite{GuillemotHPP13} showed that the approach does not directly work for approximation, and new insight seems to be necessary.

Finally, somewhat surprisingly we show that the assumption that $H$ has at least two non-edges appears to be important.
Suppose $H=K_n\setminus e$ is a complete graph on $n\geq 5$ vertices with one edge removed. 
While \Hfreecom is trivially polynomial-time solvable, due to each obstacle having only one way to be destroyed, the complexity of \Hfreedel turns out to be much more interesting.
Namely, we show that it is tightly connected to the complexity of \minhorn,
which apparently is one of the remaining open cases in the classification of the approximation complexity of CSP problems of Khanna et al.~\cite{minones}.
Hence, the following theorem shows that the case of $H$ being a complete graph without an edge may be an interesting outlier in the whole complexity picture.

\begin{theorem}\label{thm:minhorn-new1}
For any $n \ge 5$, the \kndel problem is \minhorn-complete with respect to A-reductions.
%Let $n \ge 5$.
%The \kndel problem is \minhorn-complete with respect to A-reductions,
%for instances where solutions of size greater than $\sqrt{m}$ 
%are considered infeasible.
\end{theorem}

%The reader might wonder why do we need this additional assumption about
%restricting feasible solultions in instances of \kndel in Lemma~\ref{thm:minhorn-new1}.
%This is only for technical reasons related to A-reductions.
The exact meaning of \minhorn-completeness, A-reductions and other
definitions related to the hardness of approximation for CSP problems are explained in Section~\ref{sec:minhorn}.
A direct consequence of Theorem~\ref{thm:minhorn-new1} and the work of Khanna et al.~\cite{minones} is that \kndel does not admit a $2^{\Oh(\log^{1-\epsilon} |E|)}$-approximation algorithm working in polynomial time,
for any $\epsilon>0$, where $|E|$ is the number
of edges in a given graph. 
Moreover, Theorem~\ref{thm:minhorn-new1} implies that \kndel is poly-APX-hard if and only if each \minhorn-complete problem is poly-APX-hard,
the latter being an intriguing open problem left by Khanna et al.~\cite{minones}
in their study of approximability of CSPs.

While there is no direct connection between the existence of a $\poly(\OPT)$ approximation and
poly-APX-hardness, we still believe that our reduction corroborates the hardness
of resolving approximation question of \kndel in terms of optimum value.
Intuitively, showing poly-APX-hardness should be easier than refuting
$\poly(\OPT)$ approximation. Below we state formally what our reduction actually implies.

\begin{corollary}
\label{cor:intro1}
Let $n \geq 5$. 
Then it is $\mathrm{NP}$-hard to approximate the \kndel problem 
within factor $2^{\Oh(\log^{1-\epsilon} |E|)}$ for any $\epsilon>0$, where $|E|$ is the number
of edges in a given graph. 
\end{corollary}

\begin{corollary}
\label{cor:intro2}
Let $n \geq 5$.
Then the \kndel problem admits an $n^{\delta}$-approximation for all $\delta>0$,
if and only if each \minhorn-complete problem admits an $n^{\delta_1}$-approximation for all $\delta_1>0$.
\end{corollary}

%\begin{theorem}\label{thm:minhorn}
%TODO\todo{State the min-horn result}
%\end{theorem}

%TODO 1.\todo{Should we make some table with cases? I just copied the one that was in my section, i.e. Table 1. Pawel.}

%TODO 2.\todo{If yes, should we also state complement results? It is now in Theorem~\ref{thm:main_com}. Pawel.}
\begin{comment}

\begin{table}
 \begin{center}

	\subfloat[\scriptsize \Hfreedel]{
	\begin{tabular}{|c|c|c|}%{|p{3.5cm}|p{7cm}|p{3.5cm}|}
		\hline
		$H$ & Status & Proof \\
		\hline
		$\geq  2$ non-edges & No $\poly(OPT)$  & Thm~\ref{thm:main}\\
		$K_n \setminus e$ & No $2^{\log^{1-\epsilon}(OPT)}$    & Thm~\ref{thm:horn-to-del}\\
		$K_n$ & $\Oh(1)$ & Trivial\\
		\hline
	\end{tabular}
}
%	\caption{Approximation status of \Hfreedel (\cHfreecom) with $H$ being $3$-connected.}
\subfloat[\scriptsize \Hfreecom]{
	\begin{tabular}{|c|c|c|}%{|p{3.5cm}|p{3.5cm}|p{7cm}|}
		\hline
		$H$ & Status & Proof \\
		\hline
		$\geq 2$ non-edges &No $\poly(OPT)$   & Thm~\ref{thm:main}\\
		$K_n \setminus e$ & Exact & Trivial \\
		$K_n$ & Exact & Trivial \\
		\hline
	\end{tabular}	
}
\caption{For $H$ being $3$-connected.}
%	\caption{Approximation status of \Hfreecom (\cHfreedel) with $H$ being $3$-connected.}	
\end{center}
\end{table}
\end{comment}

\paragraph*{Our techniques.} To prove our main result, Theorem~\ref{thm:main}, we employ the following strategy. 
We first consider the {\em{sandwich problem}} defined as follows: in {\sc{Sandwich $H$-Free Edge Deletion}} we are given a graph $G$ together with a subset $D$ of {\em{undeletable edges}},
and the question is whether there exists a subset $F\subseteq E(G)\setminus D$ of deletable edges for which $G-F$ is $H$-free.
Note that the sandwich problem differs from the standard \Hfreedel problem in two aspects: 
first, some edges are forbidden to be deleted, and, second, it is a decision problem about the existence of {\em{any}} solution---we do not impose any constraint on its size.
For completion, the sandwich problem is defined similarly: we have {\em{non-fillable non-edges}}, i.e., non-edges that are forbidden to be added in the solution.

The crux of the approach is to prove that {\sc{Sandwich $H$-Free Edge Deletion}} is actually NP-hard under the given assumptions on $H$.
The next step is to reduce from the sandwich problem to the standard optimization variant.
This is done by adding gadgets that emulate undeletable edges by introducing a large approximation gap, as follows.
For each undeletable edge $e$, attach a large number of copies of $H$ to $e$, so that each copy becomes an induced $H$-subgraph if $e$ gets deleted.
Then any solution that deletes the undeletable edge $e$ must have a very large cost, due to all the disjoint copies of $H$ that appear after the removal of $e$.
The assumption that $H$ is $3$-connected is very useful for showing that the constructions do not introduce any additional, unwanted copies of $H$ in the graph.

The approach for completion problems is similar. To prove Theorem~\ref{thm:main-paths-cycles} that concerns paths and cycles, we give problem-specific constructions using the same approach.
Some of them are based on previous ETH-hardness proofs for the problems, given by Drange et al.~\cite{DrangeFPV15}.

As far as Theorem~\ref{thm:minhorn-new1} is concerned, we employ a similar reduction strategy, but instead of starting from \threesat,
we start from a carefully selected \minones problem: the problem of optimizing the number of ones in a satisfying assignment to a boolean formula that uses only constraints from some
fixed family $\mathcal{F}$.
In particular, the constraint family $\mathcal{F}$ needs to be rich enough to
be \minhorn-hard, while at the same time it needs to
restrictive enough so that it can be expressed in the language of \kndel.

Our constructions are inspired by the rich toolbox of hardness proofs for kernelization and 
fixed-parameter algorithms for edge modification problems~\cite{AravindSS15,CaiC15,cai2012master,DrangeFPV15,KratschW13,GuillemotHPP13}.
In particular, the idea of considering sandwich problems can be traced back to the work of Cai and Cai~\cite{CaiC15,cai2012master}, 
who use the term {\em{quarantine}} for the optimization variants of sandwich edge modification problems, with undeletable edges and non-fillable non-edges. Quarantined problems serve a technical, auxiliary role in the work of Cai and Cai~\cite{CaiC15,cai2012master}: one first proves hardness of the quarantined problem, and then
lifts the quarantine by attaching gadgets, similarly as we do.

However, we would like to point out the new challenges that appear in the approximation setting.
Most importantly, the vast majority of previous reductions heavily use budget constraints (i.e. the fact that the solution is stipulated to be of size at most $k$) to argue the correctness;
this includes the general results of Cai and Cai~\cite{CaiC15,cai2012master}.
In our setting, we cannot use arguments about the tightness of the budget, because we need to introduce a large approximation gap at the end of the construction.
The usage of the sandwich problems {\em{without}} any budget constraints is precisely the way we overcome this difficulty.
Thus, most of the old reductions do not work directly in our setting, but of course some technical constructions and ideas can be salvaged.

\paragraph*{Outline.} In Section~\ref{sec:prelims} we introduce terminology and recall the most important facts from the previous works. 
Section~\ref{sec:main} is devoted to the proof of our main result, Theorem~\ref{thm:main}.
However, as the proof for \Hfreecom is similar to the proof for \Hfreedel,
in Section~\ref{sec:main} we present only the proof for \Hfreedel, while
the proof for \Hfreecom is postponed to Section~\ref{sec:completion}.
In Section~\ref{sec:minhorn} we discuss the proof of Theorem~\ref{thm:minhorn-new1}.
Section~\ref{sec:specific} contains the discussion of Theorem~\ref{thm:main-paths-cycles}, which is largely deferred to the appendix.
Concluding remarks and prospects on future work are in Section~\ref{sec:conc}.

%% file: preliminaries.tex
\section{Preliminaries}\label{sec:prelims}

\subsection{Basic graph definitions}

We use standard graph notation. For a graph $G$ by $V(G)$ and $E(G)$ we denote the set of vertices and edges of $G$, respectively. 
Throughout the paper we consider simple graphs only, i.e., there are no self-loops nor parallel edges.
We use $K_n$ to denote the complete graph on $n$ vertices.
By $P_\ell$ ($C_\ell$) we denote the path (cycle) with exactly $\ell$ vertices.
By $\overline{G}$ we denote the {\em{complement}} of $G$, i.e., a graph on the same vertex set, where two distinct vertices are adjacent if and
only if they were not adjacent in $G$. 
We say that a graph $G$ is {\em{H}-free}, if $G$ does not contain $H$ as an induced subgraph. 

We define a graph $G$ to be {\em{$3$-vertex-connected}} if $G$ has at least $3$ vertices, and removing
any set of at most two vertices causes $G$ to stay connected. 
For brevity, we call such graphs {\em{$3$-connected}}.

\subsection{Problems and approximation algorithms}

In the decision version the \Hfreedelcom problem, 
for a given graph $G$ and an integer $k$, 
one is to decide whether it is possible to delete (add) 
at most $k$ edges from (to) $G$ to make it $H$-free.
In particular, we consider the {\sc{$\overline P_5$-Free Deletion (Completion)}} problem, and call it {\sc{House-Free Deletion (Completion)}}.
However, in the optimization variant of \Hfreedelcom the value of $k$ is not given
and the goal is to find a minimum size solution. 
It will be clear from the context whether we refer to a decision or optimization variant.

In the {\sc{Sandwich $H$-Free Edge Deletion (Completion)}} problem we are given
a graph $G$ together with a subset $D$ of {\em{undeletable edges}} ({\em{non-fillable non-edges}}).
The question is whether there exists a subset $F\subseteq E(G)\setminus D$ ($F\subseteq \overline{E(G)}\setminus D$) of deletable (fillable) edges for which $G-F$ ($G+F$) is $H$-free. 
Note that it is a decision problem, where we ask about existence of any solution, 
i.e., we do not impose any constraint on the solution size.

%COMMENT: The Quarantineed definition will stay in Min Horn section.

Let $f$ be a fixed non-decreasing function on positive integers. 
An $f(OPT)$-{\em{factor approximation algorithm}} for a minimization problem $X$
is an algorithm that finds a solution of size at most $f(OPT) \cdot OPT$, 
where $OPT$ is the size of an optimal solution for a given instance of $X$.

%COMMENT: I wans't sure whether to use decision version of approximation or not. In intro it is optimization.

\subsection{Satisfiability and Exponential Time Hypothesis}

We employ the standard notation related to satisfiability problems. 
A 3CNF formula is a conjunction of clauses, where a clause is a disjunction of at most three literals. 
The \threesat problem asks, for a given formula $\varphi$, whether there is a satisfying assignment to $\varphi$.

The Exponential Time Hypothesis (ETH), introduced by Impagliazzo, Paturi and Zane~\cite{ImpagliazzoP01}
is now an established tool used for proving conditional lower bounds in the parameterized complexity area (see~\cite{eth-survey}
for a survey on ETH-based lower bounds). 

\begin{hypothesis}[Exponential Time Hypothesis (ETH) \cite{ImpagliazzoP01}]
There is no $2^{o(n)}$ time algorithm for \threesat, where $n$ is the number of variables of the input formula.
\end{hypothesis}

The main consequence of the Sparsification Lemma of~\cite{ImpagliazzoP01}
is the following theorem: there is no subexponential algorithm for \threesat
even in terms of the number of clauses of the formula.

\begin{theorem}[\cite{ImpagliazzoP01}]
Unless ETH fails, there is no $2^{o(n+m)}$ time algorithm for \threesat, where $n$, $m$ are the number of variables, and clauses, respectively. 
\end{theorem}

%% file: reduction.tex
\newcommand{\vars}{\mathtt{vars}}
\newcommand{\cls}{\mathtt{cls}}

\section{Hardness for $3$-connected $H$}\label{sec:main}

In this section we present the proof of Theorem~\ref{thm:main}
for \Hfreedel, while a similar proof for \Hfreecom
is deferred to Section~\ref{sec:completion}.

\subsection{Deletion problems}\label{sec:deletion}

We start with proving hardness of the sandwich problem.

\begin{lemma}\label{thm:3sat-3conQuar}
	Let $H$ be a $3$-connected graph with at least $2$ non-edges.
	There is a polynomial-time reduction, which given an instance of \threesat with $n$ variables and $m$ clauses, creates an equivalent instance of \gHfreedel
	with $\Oh(n+m)$ edges. Consequently, \gHfreedel is $\mathrm{NP}$-hard for such graphs $H$.
\end{lemma}

\begin{proof}
        Let $\varphi$ be the given formula in 3CNF, and let $\vars$ and $\cls$ be the sets of variables and clauses of $\varphi$.
	By standard modifications of the formula, we may assume that each clause contains exactly three literals of pairwise different variables. 
	We construct an instance $G$ of \gHfreedel as follows.
	The graph $G$ is created from three types of gadgets: a clause gadget, a variable gadget, and a connector gadget. 
	They are depicted in Figure~\ref{fig:H-del-gadgets}, where presented edges are deletable, and all others are undeletable.
	
	We first explain constructions of the gadgets, and then discuss connections between them. 
	For each variable $x \in \vars$, we create a variable gadget $G^x$, which is the graph $H$ with two added edges $e_x$ and $e_{\nn{x}}$ in place of any two non-edges of $H$. 
	In the graph $H_x$, all edges are marked as undeletable except $e_x$ and $e_{\nn{x}}$. 
	Intuitively, deletion of the edge $e_x$ or $e_{\nn{x}}$ mimics an assignment of the corresponding literal to true.
	The variable gadget forbids simultaneous assignments of both literals to true. If we delete both edges $e_x$
	and $e_{\nn{x}}$, we get an induced subgraph $H$ in which we cannot delete any edge.  
	
	Each clause $c = \ell_1 \vee \ell_2 \vee \ell_3 \in \cls$ has the corresponding clause gadget $H^{c}$, which is a copy of the graph $H$. As $H^c$ is $3$-connected,
	it has at least $3$ edges. We pick arbitrarily three edges of $H^c$ and label them by $e_{\ell_1}, e_{\ell_2}, e_{\ell_3}$. We mark all others edges as undeletable. 
	In order to make the clause gadget $H$-free, we have to delete at least one edge from $e_{\ell_1}, e_{\ell_2}, e_{\ell_3}$ 
  (note that some of the three distinguished edges might potentially share an endpoint). 
	Intuitively, deletion of the edge labeled by $e_\ell$ corresponds to assigning value true to literal~$\ell$. 
	
	The third type of gadgets is the connector gadget. The connector gadget $C$ is a copy of the graph $H$, with one added edge in place of any non-edge of $H$. We label this edge as $e_{in}$. 
	In $C$, there also exists another edge that does not share any of its endpoints with $e_{in}$. 
	To see this, for the sake of contradiction suppose that every edge of $C$ is incident to one of the endpoints of $e_{in}$.
	If $C$ has at least two vertices other than these endpoints, then the endpoints of $e_{in}$ form a vertex cut of size $2$ separating them, a contradiction with $3$-connectedness of $H$.
	Otherwise $C$ has only one vertex other than the endpoints of $e_{in}$, so $H$ has at most $3$ vertices; again, a contradiction with the $3$-connectedness of $H$,
  as we assume $H$ to have at least $2$ non-edges.
	We select any edge in $H$ that does not share endpoints with $e_{in}$, and we label it as $e_{out}$.
	Edges $e_{in}$ and $e_{out}$ are made deletable, and all other edges of $C$ are made undeletable. 
	Note that deletion of the edge $e_{in}$ creates an induced subgraph $H$, and then we have to delete $e_{out}$ in order to destroy this subgraph.
	
		\begin{figure}
	  \centering
	  \subfloat[Variable gadget $G^x$]{
	      \raisebox{0cm}{\def\svgwidth{0.26\textwidth}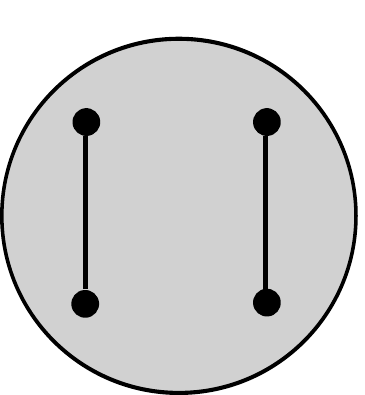}
	  }
	  \quad
	  \subfloat[Clause gadget $H^c$]{
	      \raisebox{0cm}{\def\svgwidth{0.26\textwidth}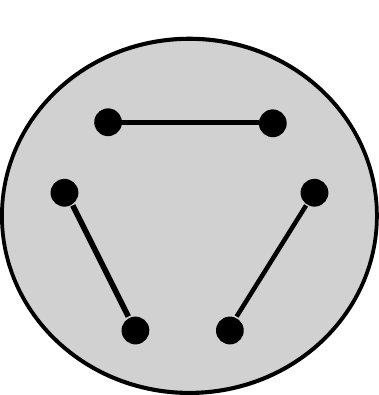}
	  }
	  \quad
	  \subfloat[Connector gadget $C$]{
	      \raisebox{0cm}{\def\svgwidth{0.23\textwidth}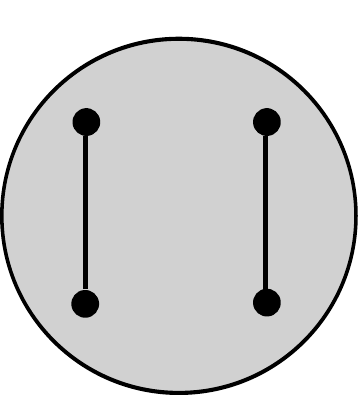}
	  }
	  \caption{Gadgets for \gHfreedel.}\label{fig:H-del-gadgets}
       \end{figure}
	
	Knowing the structure of all gadgets, we can proceed with the main construction of our reduction. 
	
	%Our goal is to convert a given $3$-CNF formula $F$ into an instance of parameterized Guaranteed $H$-Deletion such that a satisfiable formula corresponds to YES instance of  parameterized Guaranteed $H$-Deletion problem. We set the parameter $k$ equal to $3 (|V(H)|+3) m$ where m is a number of clauses in the given formula $F$.
	
	Given a formula $\varphi$, for each clause $c \in \cls$ and variable $x \in \vars$, we create the clause gadget $H^c$ and the variable gadget $G^x$, respectively. 
	Moreover, for each literal $\ell$ belonging to the clause $c \in \cls$, 
	we create a chain $C_1^{\ell,c}, C_2^{\ell,c}, \ldots, C_{p + 2}^{\ell,c}$ consisting of $p + 2$ copies of the connector gadget, where $p=|V(H)|$.
	This chain is constructed in the following way:
	the edge $e_{out}$ of $C_i^{\ell,c}$ is identified with the edge $e_{in}$ of $C_{i+1}^{\ell,c}$, for $i = 1, \ldots ,p+1$. 
	We also identify the edge $e_{out}$ in the subgraph $C_{p + 2}^{\ell,c}$ with the edge 
	$e_{\ell}$ in the variable gadget of the variable of $\ell$. 
	Moreover, the edge $e_{in}$ in the subgraph $C_1^{\ell,c}$ is identified with the edge $e_{\ell}$ from the clause gadget 
	$H^c$. We use those chains to not allow the copy of $H$ to be shared by any two gadgets, and we will prove it in the claim below.
	
	Clearly, the constructed graph $G$ has at most $\Oh(n+m)$ edges.
	
	%Finally, we set the parameter $k$ equal to $3 (|V(H)|+2) m$ proceed with the proof of the equivalnce of the instances. 
	
	\begin{claim}\label{cl:Hdel-there}
	  If $G$ is a YES instance, then $\varphi$ is satisfiable. 
	\end{claim}
	\begin{proof}
	Take any solution to the instance $G$.
	Note that in each clause gadget we must delete at least one edge. We set the literals corresponding to the deleted edges to true, thus satisfying every clause. 
	We prove now that for each variable $x$ we have not set both literals $x$ and $\neg{x}$ to true, so that we can find a true/false assignment to the variables that sets the literals accordingly. 
	Deletion of an edge in the clause gadget propagates deletions 
	up to the variable gadget via the chain of connector gadgets. This happens because the deletion of $e_{in}$ in $C_1^{\ell,c}$ forces us to delete the $e_{out}$ in $C_1^{\ell,c}$, 
	which is $e_{in}$ in $C_2^{\ell,c}$, so we are forced to delete $e_{out}$ in $C_2^{\ell,c}$, and so on. Following the chain of connector gadgets, 
	it is easy to see that the edge $e_\ell$ must be deleted in the corresponding variable gadget. As the solution to the instance $G$ cannot delete both edges $e_x$ and $e_{\neg{x}}$ in 
	any variable gadget at the same time, we obtain that there are no variables with both of its literals set to true. 
	\cqed\end{proof}
	
	\begin{claim}\label{cl:Hdel-back}
	 If $\varphi$ is satisfiable, then $G$ is a YES instance.
	\end{claim}
	\begin{proof} Consider a true/false assignment that satisfies the formula $\varphi$ and delete all edges in all clause gadgets that correspond to literals taking value true. 
	Propagate deletions to all the connector and variable gadgets, as in the proof of Claim~\ref{cl:Hdel-there}.
	It remains to prove that the obtained graph is indeed an $H$-free graph. By counting the number of edges in each gadgets, it follows that after the deletions, 
	all gadgets become not isomorphic to $H$: in every variable gadget, we deleted exactly one edge, in every clause gadget, 
	we deleted at least one edge, and in each connector gadget we deleted zero or two edges. 
	So if the obtained graph contains an induced subgraph of $H$, then  $H$ is distributed across several gadgets. 
	However, this is also not possible for the following reason. 
	
	For the sake of contradiction, suppose after the deletions there is an induced copy $H'$ of the graph $H$. 
	Since $H'$ is connected and is distributed among more than one gadget,
	there have to be two different gadgets $G_1, G_2$ that share a vertex, for which $H'$ contains both some vertex $u\in V(G_1)\setminus V(G_2)$, and some vertex $v\in V(G_2)\setminus V(G_1)$.
	Since $H'$ is $3$-connected, there are $3$ internally vertex-disjoint paths in $H'$ that lead from $u$ to $v$.
	But every two gadgets share at most two common vertices, so at least one of these paths, say $P$, avoids $V(G_1)\cap V(G_2)$. 
	Since the path $P$ avoids $V(G_1)\cap V(G_2)$, from the construction of $G$ it easily follows 
	that such path $P$ contains at least one vertex of some variable gadget 
  and at least one vertex of some clause gadget.
	However, the distance between $e_{in}$ and $e_{out}$ in each connector gadget is at least $1$, 
  so the distance between any variable gadget and any clause gadget is at least $|V(H)|$.
	But the path $P$ is entirely contained in $H'$, thus its length is at most $|V(H)|-1$, a contradiction.
	\cqed\end{proof}  
	
	Claims~\ref{cl:Hdel-there} and~\ref{cl:Hdel-back} ensure that the output instance $G$ is equivalent to the input instance $\varphi$ of \threesat, so we are done.
	\end{proof}

	Now, we show how to reduce \gHfreedel to the optimization variant of \Hfreedel. 
	Note that we only require $H$ to have at least one non-edge; this is because we will reuse this lemma in the next section.
	
	\begin{lemma}\label{thm:Quar-3con}
	Let $H$ be a $3$-connected graph with at least one non-edge, and $p(\cdot)$ be a polynomial with $p(\ell) \geq \ell$, for all positive $\ell$. Then there is a polynomial-time reduction which, given 
	an instance $G$ of \gHfreedel, creates an instance $(G',k)$ of \Hfreedel, such that: 
	\begin{itemize}
	 \item $k$ is the number of deletable edges in $G$;
	 \item $G'$ has $\Oh(p(k) \cdot |E(G)| \cdot |E(H)|)$ edges;
	 \item If $G$ is a YES instance, then $(G',k)$ is a YES instance;
	 \item If $G$ is a NO instance, then $(G', p(k))$ is a NO instance.
	\end{itemize}
	%Let $H$ be a $3$-connected graph with at least $2$ non-edges. There is a polynomial-time reduction, which given an instance of \gHfreedel, 
	%creates an instance $(G,k)$ of \Hfreedel, such that $k = \Oh(n+m)$.
	\end{lemma}
	\begin{proof}

	We create $G'$ in the following way. %If our instance do not contain undeletable edges then we immideately %output YES as we can simply delete all edges. 
	For each undeletable edge $uv$, we add $p(k)$ copies $H_i^{uv}$ of the graph $H$, $i=1,\ldots,p(k)$. In each copy, 
	we choose any non-edge $u_iv_i$ and identify the vertex $u_i$ with $u$, and $v_i$ with $v$. 
	The construction is presented in Figure~\ref{fig:H-del-forbid}.
	
	 \begin{figure}
	  \centering
	      \raisebox{0cm}{\def\svgwidth{0.26\textwidth}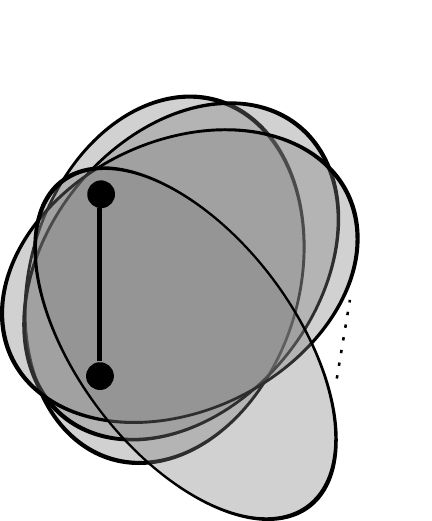}
	  
	  \caption{Gadgets $H^{uv}_i$ for \Hfreedel.}\label{fig:H-del-forbid}
       \end{figure}

	Note that if we delete the edge $uv$ in $G'$, we also must delete at least one edge in every $H_i^{uv}$. Hence, at least $p(k)+1$ edges will be deleted in such a situation.
	With this observation in mind, we proceed to the proof of the correctness.
	
	\begin{claim}\label{cl:Hdel-quar-there}
	If $G$ is a YES instance, then $(G',k)$ is a YES instance.
	\end{claim}
	\begin{proof}
	Let $F$ be a subset deletable edges, such that $G-F$ is $H$-free. Obviously $|F|\leq k$, because there are $k$ deletable edges in $G$ in total.
	We will prove that $G'-F$ is also $H$-free, which implies that $(G',k)$ is a YES instance.
	
	Let us assume otherwise, that there is an induced copy $H'$ of $H$ in $G'$. 
	Since $G-F$ is $H$-free, we have that $H'$ has to contain at least one vertex of $V(G')\setminus V(G)$.
	Say that $H'$ contains some vertex $x$ of $V(H_i^{uv})\setminus V(G)$, for some undeletable edge $uv$ and some index $i$.
	The edge $uv$ is undeletable in $G$, so it is not included in $F$.
	Consequently, the subgraph of $G'$ induced by $V(H_i^{uv})$ contains one more edge than $H$, so it is not isomorphic to $H$.
	We conclude that $H'$ must contain some vertex $y$ that lies outside of $V(H_i^{uv})$.
	Since $H$ is $3$-connected, there are $3$ internally vertex-disjoint paths between $x$ and $y$ in $H$.
	However, in $G$, the set $V(H_i^{uv})\cap V(G)=\{u,v\}$ is a vertex cut of size $2$ that separates $x$ and $y$.
	This is a contradiction, so $G'-F$ is indeed $H$-free.
	\cqed\end{proof}
	
	\begin{claim}\label{cl:Hdel-quar-back}
	 If $G$ is a NO instance, then $(G', p(k))$ is a NO instance.
	\end{claim}
	\begin{proof}
	For the sake of contradiction, suppose there is a set $F'$ of at most $p(k)$ edges of $G'$, such that $G'-F'$ is $H$-free. 
	Note that, $F'$ has to contain at least one undeletable edge $uv$, as otherwise $F' \cap E(G)$ would be a solution to $G$. 
	But then $F'$ has to contain at least $p(k)$ more edges inside gadgets $H_i^{uv}$, for $i=1,2,\ldots,p(k)$, which is a contradiction with $|F'|\leq p(k)$.
	\cqed\end{proof}
	
	Claims~\ref{cl:Hdel-quar-there} and~\ref{cl:Hdel-quar-back} ensure the correctness of the reduction, and hence we are done.
	\end{proof}
	
	By composing the reductions of Lemmas~\ref{thm:3sat-3conQuar} and~\ref{thm:Quar-3con}, we can deduce the part of Theorem~\ref{thm:main} concerning deletion problems.
	Indeed, suppose \Hfreedel admitted a polynomial-time $q(\OPT)$-factor approximation algorithm, for some polynomial $q$.
	Take any instance of \threesat, and apply first the reduction of Lemma~\ref{thm:3sat-3conQuar}, 
	and then the reduction of Lemma~\ref{thm:Quar-3con} for polynomial $p(\ell)=q(\ell)\cdot \ell+1$.
	Finally, observe that the application of the hypothetical approximation algorithm for \Hfreedel to the resulting instance would
	resolve whether the optimum value is at most $k$ or at least $p(k)$, which, by Lemma~\ref{thm:Quar-3con}, resolves whether the input instance of \threesat is satisfiable.
	The subexponential hardness of approximation under ETH follows from the same reasoning and the observation that the value of $k$ in the output instance
	is bounded linearly in the size of the input formula.
	
\input minhorn

%% file: Hdel_variable.pdf_tex
%% Creator: Inkscape inkscape 0.48.3.1, www.inkscape.org
%% Accompanies image file 'Hdel_variable.pdf' (pdf, eps, ps)
\begingroup%
  \makeatletter%
  \providecommand\color[2][]{%
    \errmessage{(Inkscape) Color is used for the text in Inkscape, but the package 'color.sty' is not loaded}%
    \renewcommand\color[2][]{}%
  }%
  \providecommand\transparent[1]{%
    \errmessage{(Inkscape) Transparency is used (non-zero) for the text in Inkscape, but the package 'transparent.sty' is not loaded}%
    \renewcommand\transparent[1]{}%
  }%
  \providecommand\rotatebox[2]{#2}%
  \ifx\svgwidth\undefined%
    \setlength{\unitlength}{105.00719736bp}%
    \ifx\svgscale\undefined%
      \relax%
    \else%
      \setlength{\unitlength}{\unitlength * \real{\svgscale}}%
    \fi%
  \else%
    \setlength{\unitlength}{\svgwidth}%
  \fi%
  \global\let\svgwidth\undefined%
  \global\let\svgscale\undefined%
  \makeatother%
  \begin{picture}(1,1.08325343)%
     \put(0,0){\includegraphics[width=\unitlength]{Hdel_variable.pdf}}%
    \put(0.27161558,0.48172082){\color[rgb]{0,0,0}\makebox(0,0)[lb]{\smash{{\tn{$e_x$}}}}}%
    \put(0.5763566,0.4817208){\color[rgb]{0,0,0}\makebox(0,0)[lb]{\smash{{\tn{$e_{\neg x}$}}}}}%
    \put(0.30970819,1.01501769){\color[rgb]{0,0,0}\makebox(0,0)[lb]{\smash{{\tn{$H \cup e_{x} \cup {e_{\neg x}}$}}}}}%
  \end{picture}%
\endgroup%

%% file: Hdel_clause.pdf_tex
%% Creator: Inkscape inkscape 0.48.3.1, www.inkscape.org
%% Accompanies image file 'Hdel_clause.pdf' (pdf, eps, ps)
\begingroup%
  \makeatletter%
  \providecommand\color[2][]{%
    \errmessage{(Inkscape) Color is used for the text in Inkscape, but the package 'color.sty' is not loaded}%
    \renewcommand\color[2][]{}%
  }%
  \providecommand\transparent[1]{%
    \errmessage{(Inkscape) Transparency is used (non-zero) for the text in Inkscape, but the package 'transparent.sty' is not loaded}%
    \renewcommand\transparent[1]{}%
  }%
  \providecommand\rotatebox[2]{#2}%
  \ifx\svgwidth\undefined%
    \setlength{\unitlength}{109.10454303bp}%
    \ifx\svgscale\undefined%
      \relax%
    \else%
      \setlength{\unitlength}{\unitlength * \real{\svgscale}}%
    \fi%
  \else%
    \setlength{\unitlength}{\svgwidth}%
  \fi%
  \global\let\svgwidth\undefined%
  \global\let\svgscale\undefined%
  \makeatother%
  \begin{picture}(1,1.04257259)%
    \put(0,0){\includegraphics[width=\unitlength]{Hdel_clause.pdf}}%
    \put(0.28002749,0.3903059){\color[rgb]{0,0,0}\makebox(0,0)[lb]{\smash{{\tn{$e_{l_2}$}}}}}%
    \put(0.60998625,0.3903059){\color[rgb]{0,0,0}\makebox(0,0)[lb]{\smash{{\tn{$e_{l_3}$}}}}}%
    \put(0.48533517,0.778924){\color[rgb]{0,0,0}\makebox(0,0)[lb]{\smash{{\tn{$e_{l_1}$}}}}}%
    \put(0.46333792,0.9768994){\color[rgb]{0,0,0}\makebox(0,0)[lb]{\smash{{\tn{$H$}}}}}%
  \end{picture}%
\endgroup%

%% file: Hdel_connector.pdf_tex
%% Creator: Inkscape inkscape 0.48.4, www.inkscape.org
%% Accompanies image file 'Hdel_connector.pdf' (pdf, eps, ps)
\begingroup%
  \makeatletter%
  \providecommand\color[2][]{%
    \errmessage{(Inkscape) Color is used for the text in Inkscape, but the package 'color.sty' is not loaded}%
    \renewcommand\color[2][]{}%
  }%
  \providecommand\transparent[1]{%
    \errmessage{(Inkscape) Transparency is used (non-zero) for the text in Inkscape, but the package 'transparent.sty' is not loaded}%
    \renewcommand\transparent[1]{}%
  }%
  \providecommand\rotatebox[2]{#2}%
  \ifx\svgwidth\undefined%
    \setlength{\unitlength}{103.04318101bp}%
    \ifx\svgscale\undefined%
      \relax%
    \else%
      \setlength{\unitlength}{\unitlength * \real{\svgscale}}%
    \fi%
  \else%
    \setlength{\unitlength}{\svgwidth}%
  \fi%
  \global\let\svgwidth\undefined%
  \global\let\svgscale\undefined%
  \makeatother%
  \begin{picture}(1,1.10390038)%
    \put(0,0){\includegraphics[width=\unitlength]{Hdel_connector.pdf}}%
    \put(0.2612651,0.49090248){\color[rgb]{0,0,0}\makebox(0,0)[lb]{\smash{{\tn{$e_{in}$}}}}}%
    \put(0.34666626,1.03436406){\color[rgb]{0,0,0}\makebox(0,0)[lb]{\smash{{\tn{$H \cup e_{in}$}}}}}%
    \put(0.55628719,0.49090248){\color[rgb]{0,0,0}\makebox(0,0)[lb]{\smash{{\tn{$e_{out}$}}}}}%
  \end{picture}%
\endgroup%

%% file: Hdel_forbid.pdf_tex
%% Creator: Inkscape inkscape 0.48.3.1, www.inkscape.org
%% Accompanies image file 'Hdel_forbid.pdf' (pdf, eps, ps)
\begingroup%
  \makeatletter%
  \providecommand\color[2][]{%
    \errmessage{(Inkscape) Color is used for the text in Inkscape, but the package 'color.sty' is not loaded}%
    \renewcommand\color[2][]{}%
  }%
  \providecommand\transparent[1]{%
    \errmessage{(Inkscape) Transparency is used (non-zero) for the text in Inkscape, but the package 'transparent.sty' is not loaded}%
    \renewcommand\transparent[1]{}%
  }%
  \providecommand\rotatebox[2]{#2}%
  \ifx\svgwidth\undefined%
    \setlength{\unitlength}{124.23507717bp}%
    \ifx\svgscale\undefined%
      \relax%
    \else%
      \setlength{\unitlength}{\unitlength * \real{\svgscale}}%
    \fi%
  \else%
    \setlength{\unitlength}{\svgwidth}%
  \fi%
  \global\let\svgwidth\undefined%
  \global\let\svgscale\undefined%
  \makeatother%
  \begin{picture}(1,1.20770099)%
    \put(0,0){\includegraphics[width=\unitlength]{Hdel_forbid.pdf}}%
    \put(0.25093476,0.54472209){\color[rgb]{0,0,0}\makebox(0,0)[lb]{\smash{{\tn{$uv$}}}}}%
    \put(0.32820762,1.15002611){\color[rgb]{0,0,0}\makebox(0,0)[lb]{\smash{{\tn{$p(k) \times (H \cup uv)$}}}}}%
    \put(0.3410866,1.01479854){\color[rgb]{0,0,0}\makebox(0,0)[lb]{\smash{{\tn{$H^{uv}_1$}}}}}%
    \put(0.56002684,1.00192004){\color[rgb]{0,0,0}\makebox(0,0)[lb]{\smash{{\tn{$H^{uv}_2$}}}}}%
    \put(0.77896598,0.83449542){\color[rgb]{0,0,0}\makebox(0,0)[lb]{\smash{{\tn{$H^{uv}_3$}}}}}%
    \put(0.79828413,0.14547921){\color[rgb]{0,0,0}\makebox(0,0)[lb]{\smash{{\tn{$H^{uv}_{p(k)}$}}}}}%
  \end{picture}%
\endgroup%

%% file: minhorn.tex
\newcommand{\nvars}{n_{\mathrm{vars}}}

\section{Connections with \minhorn}\label{sec:minhorn}

In this section we prove Theorem~\ref{thm:minhorn-new1}.
First, we need to introduce some definitions and notation regarding 
\minhorn hardness and completeness.

Khanna et al.~\cite{minones} attempted to establish a full classification 
of approximability of boolean constraint satisfaction problems.
In particular, many problems have been classified as APX-complete or poly-APX-complete.
Even though some cases remained unresolved,
Khanna et al.~\cite{minones} grouped them into classes,
such that all problems from the same class are equivalent
(with respect to appropriately defined reductions)
to a particular representative problem.
One such representative problem is \minhorn, defined as follows:
Given is a boolean formula $\varphi$ in CNF that contains only unary clauses, and clauses with three literals out of which exactly one is negative.
The problem asks for minimizing the number of ones in a satisfying assignment for $\varphi$.

We are not going to operate on instances of \minhorn directly, so the definition above is given only in order to complete the picture for the reader.
Instead, we will rely on the approximation hardness results exhibited by Khanna et al.~\cite{minones}, which relate the approximability of various
boolean CSPs to \minhorn. In particular, it is known that \minhorn does not admit a $2^{\Oh(\log^{1-\epsilon} \nvars)}$ approximation algorithm, unless $\mathrm{P}=\mathrm{NP}$,
where $\nvars$ is the number of variables in the instance.
On the other hand, it is an open problem whether any \minhorn-complete
problem (under $A$-reductions, defined below) is actually poly-APX-complete.

\begin{comment}
As we are not going to operate on instances of \minhorn,
we decided not to define this problem here and ask the
reader to treat \minhorn as a hard problem complete in its class\footnote{The
definition of \minhorn together with a thorough description of
related issues can be found in~\cite{minones}.}.
However, we need to define the exact notion of reductions
used in the definition of the class of \minhorn-complete
problems, namely the $A$-reduction.
\end{comment}

\begin{definition}[A-reducibility, Definition 2.6 of \cite{minones}]
A combinatorial optimization problem is said to be an NPO
problem if instances and solutions can be recognized in polynomial time,
solutions are polynomially-bounded in the input size, and the objective
function can be computed in polynomial time from an instance and a solution.

An NPO problem $P$ is said to be {\em A-reducible} to an NPO problem $Q$,
denoted $P \le_A Q$, if there are two polynomial-time computable functions $F$ and $G$
and a constant $\alpha$, such that:
\begin{enumerate}
  \item For any instance $\mathcal{I}$ of $P$, $F(\mathcal{I})$ is an instance of $Q$.
  \item For any instance $\mathcal{I}$ of $P$ and any feasible solution $\mathcal{S}'$
  for $F(\mathcal{I})$, $G(\mathcal{I}, \mathcal{S}')$ is a feasible solution for $\mathcal{I}$.
  \item For any instance $\mathcal{I}$ of $P$ and any $r \ge 1$, if $\mathcal{S}'$ is an $r$-approximate
  solution for $F(\mathcal{I})$, then $G(\mathcal{I}, \mathcal{S}')$ is an $(\alpha r)$-approximate
  solution for $\mathcal{I}$.
\end{enumerate}
\end{definition}

Intuitively, $A$-reductions preserve approximability problems up to a constant factor (or higher).
As a source of \minhorn-hardness we will use the \minones problem, defined below,
for a particular choice of the family of constraints $\mathcal{F}$.

In the \minones problem, we are given a ground set of boolean variables $X$
together with a set of boolean constraints. 
Each constraint $f$ is taken from a specified family $\mathcal{F}$, and $f$ is applied to some tuple of variables from $X$.
The goal of the problem is to find an assignment satisfying all the constraints,
while minimizing the number of variables set to one.
Note that the family $\mathcal{F}$ is considered a part of the problem definition, not part of the input.
In order to use known results for the \minones problem we need
to define some properties of boolean constraints.
\begin{itemize}
	\item A boolean constraint $f$ is called \emph{weakly positive} if it can be expressed using a CNF formula that has at most one negated variable in each clause. 
	\item A boolean constraint $f$ is \emph{$0$-valid} if the all-zeroes assignment satisfies it. 
	%\item A constraint is \emph{affine} if it can be expressed as a conjunction of linear equalities over $\mathbb{Z}_2$. 
%	\item A constraint is \emph{2cnf} if it is expressible as a $2$CNF formula. 
	\item  A boolean constraint $f$ is IHS-$B^+$ if it can be expressed using a CNF formula in which the clauses are all of one of the following types: 
	$x_1\vee \dots \vee x_k$ for some positive integer $k\leq B$, or $\neg{x_1}\vee x_2$, or $\neg{x_1}$. IHS-$B^-$ constraints are defined analogously, with every literal being replaced by its complement. 
\end{itemize}

%Let us start with introducting two auxiliary problems, namely \minones and \quar,
%which are only used in this section and for this reason we define them only now.

The definition can be naturally extended to families of constraints, e.g., a family of constraints is weakly positive if all its constraints are weakly positive. 
We say that a family of constraints is IHS-$B$ if it is either IHS-$B^+$ or IHS-$B^-$ (or both).
The following result was proved by Khanna et al.~\cite{minones}.

\begin{theorem}[Lemmas 8.7 and 8.14 from \cite{minones}]
\label{thm:cited}
If a family of constraints $\mathcal{F}$ is weakly positive, but it is neither $0$-valid nor IHS-$B$ for any constant $B$,
then the problem \minones is \minhorn-complete under $A$-reductions; that is, there is an $A$-reduction from \minhorn to \minones
and an $A$-reduction from \minones to \minhorn.
Consequently, it is NP-hard to approximate \minones within factor $2^{\Oh(\log^{1-\epsilon}\nvars)}$ for any $\epsilon>0$,
where $\nvars$ is the number of variables in the given instance.
\end{theorem}

% From now on we think that set of constraints  $\mathcal{F}$ is weakly positive but not $0$-valid, 2cnf, IHS-B, affine.

Our strategy for the proof of Theorem~\ref{thm:minhorn-new1} is as follows.
In Section~\ref{ssec:minhorn1} we show a reduction from \minones
to a properly defined quarantined version of \kndel.
Next, in Section~\ref{ssec:minhorn2} we show a reduction which removes the quarantine.
Finally, in Section~\ref{ssec:minhorn3} we conclude the proof of Theorem~\ref{thm:minhorn-new1}
and show the completeness with respect to $A$-reductions.

Note that having Theorem~\ref{thm:minhorn-new1}, we can immediately
infer Corollaries~\ref{cor:intro1},\ref{cor:intro2} using Theorem~\ref{thm:cited} and the definition of an $A$-reduction.

\subsection{From \minones to \quar}
\label{ssec:minhorn1}

In the \quar problem we are given a graph $G$,
  some edges of which are marked as undeletable.
\quar is an optimization problem,
 where the goal is to obtain an $H$-free graph by removing the minimum number of deletable edges.

Next, we define the family of constraints that will be used in the \minones problem.

\begin{definition}
We define the following constraints:
\begin{itemize}
  \item a constraint $f_1(x_1,x_2,x_3)$, which is equal to zero if and only if exactly one of the variables $x_1, x_2, x_3$ is set to $1$;
  \item a constraint $f_2(x)=x$.
 % \item a constraint $f_3(x)=\neg{x}$.
\end{itemize}
The family of constraints $\mathcal{F'}$ is defined as $\mathcal{F'}=\{f_1, f_2\}$.
\end{definition}

A direct check, presented below, verifies that $\mathcal{F'}$ has the properties needed to claim, using Theorem~\ref{thm:cited}, that \minoness is \minhorn-hard.

\begin{lemma}
\label{lem:fproperties}
The family of constraints $\mathcal{F'}=\{f_1,f_2\}$ is weakly positive, and at the same time 
it is neither $0$-valid, nor IHS-$B$ for any $B$.
\end{lemma}

\begin{proof}
Note that $f_1$ is weakly positive since $f_1(x_1,x_2,x_3)=(\neg{x_1} \vee x_2 \vee x_3) \wedge (x_1 \vee \neg{x_2} \vee x_3) \wedge (x_1 \vee x_2 \vee \neg{x_3})$.
Constraint $f_2$ is clearly weakly positive by definition.
As $f_2$ is not $0$-valid, we have that $\mathcal{F'}$ is not $0$-valid either. 

We prove now that $f_1$ is not IHS-$B$ for any $B$. 
First, observe that any CNF formula expressing $f_1$ cannot contain 
a clause with only positive literals,
as such a clause would not be satisfied 
by the assignment $x_1=x_2=x_3=0$, which in turn satisfies $f_1$.
Similarly, no clause can have only negative literals.
Due to the definition of IHS-$B$, the only remaining case
is a $2$-clause with one positive and one negative literal.
Without loss of generality, consider a clause
$x_1\vee \neg{x_2}$. Observe, that it is not satisfied
by the assignment $x_1=0,\, x_2=x_3=1$, which however satisfies $f_1$.
Therefore $f_1$, and consequently $\mathcal{F'}$, is not IHS-$B$ for any B.
\end{proof}

Consequently, Theorem~\ref{thm:cited} and Lemma~\ref{lem:fproperties} together imply that \minoness is \minhorn-hard under $A$-reductions. We now give our main reduction, from \minoness to \quarkndel.

\begin{lemma}\label{lem:horn-red1}
	Let $n \geq 5$. There is a polynomial-time computable transformation $T$ which, given
  an instance $\mathcal{I}$ of the \minoness problem, outputs an instance $T(\mathcal{I})$
  of the \quarkndel problem, such that:
  \begin{itemize}
    \item if $\mathcal{I}$ admits a satisfying assignment with $k$ ones,
    then there is a solution of cost $\Delta \cdot k$ for the instance~$T(\mathcal{I})$,
    \item if $T(\mathcal{I})$ admits a solution of cost $k'$, 
    then there is a satisfying assignment with $\lfloor k' / \Delta \rfloor$ ones for the instance~$\mathcal{I}$,
  \end{itemize}
  where $\Delta = 9\nvars^2+2$ and $\nvars$ is the number of variables in $\mathcal{I}$.
\end{lemma}

\begin{figure}
	\centering
	{
		\def\svgwidth{0.6\textwidth}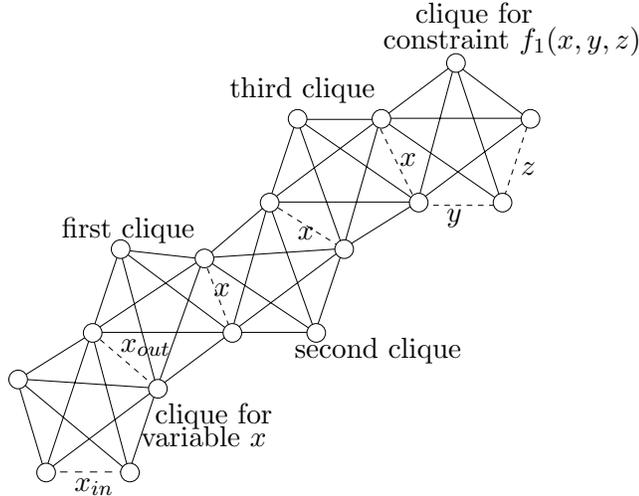
	}
	\caption{Gadgets for {\sc{Quarantined $K_n\setminus e$-Free Edge Deletion}}. Deletable edges are shown by dashed lines.}\label{fig:kndel}
\end{figure}

\begin{proof}
	First, we show how to transform an instance $\mathcal{I}$ (with a formula $\varphi$) of \minoness into an instance $T(\mathcal{I})$ (with a graph $G$) of  \quarkndel. 
	Given an instance $\mathcal{I}$, for any constraint $f_1(x,y,z)$ we create a separate clique $K_n$, which will be called the {\em{constraint clique}}. 
	We arbitrarily choose three edges in the clique and label them $x, y, z$. 
	Mark all edges as undeletable except edges labelled by $x, y, z$. 
	Moreover,  for each variable $x$ we additionally create a clique $K_n$ (called further the {\em{variable clique}}), 
	and mark all edges in the clique as undeletable except two edges, which we label by $x_{in}, x_{out}$. 
	The edges $x_{in}, x_{out}$ are selected arbitrarily, however we require that they do not share common endpoints. 
%	If there is a clause $f_2(x)=x$ in the instance $\mathcal{I}$, then we delete edge labelled by $x_{in}$ in the corresponding variable clique. 
	%If there is a clause $f_3(x)=\neg{x}$ in the instance $\mathcal{I}$, then we mark the edge $x_{in}$ in the corresponding variable clique as undeletable. 
	%Note that if both $x$ and $\neg{x}$ are present in $\mathcal{I}$ then the instance is not satisfiable. 

	Now we connect the variable cliques with the constraint cliques. 
	For each variable $x$ and a constraint $f_1$ of the instance $\mathcal{I}$ which contains $x$ among its arguments, we add three cliques, as shown in Figure~\ref{fig:kndel},
	such that the following properties are satisfied:
	\begin{itemize}
		\item The first added clique shares with the variable clique of $x$ only the edge $x_{out}$.
		\item The second added clique shares one deletable edge with the first clique and a different deletable edge with the third clique. 
		      Label both these deletable edges by $x$.
		\item The third added clique shares with the clique corresponding to the constraint only the edge labelled (in the constraint clique) by $x$.
	\end{itemize}
	All the other edges of the introduced cliques, not mentioned above, are marked as undeletable. Note that each of the introduced cliques shares two edges with two different cliques.
	We may perform this construction so that these two edges never share endpoints (as depicted Figure~\ref{fig:kndel}), and hence we will assume this property.
	
	Denote by $\delta(x)$ the number of occurrences of the variable $x$ in all $f_1$-type constraints. 
	Note that, by removing superfluous copies of the same constraint, we can assume that all $f_1$-type constraints are pairwise different, so in particular there is at most $\nvars^3$ of them.
	As each variable can occur in one constraint at most three times, for any variable $x$ we have $\delta(x) \le 3\nvars^2$.
	
	Next, for each variable $x$ we add $3\cdot(3\nvars^2-\delta(x))$ or $3\cdot (3\nvars^2-\delta(x))+1$  cliques that share the deletable edge $x_{in}$ from the variable clique of $x$,
	and are otherwise disjoint.
	Moreover, in each such clique we make one more edge deletable; we label it by $x$. 
	We add $3\cdot (3\nvars^2-\delta(x))$ cliques if the formula does contain the clause $f_2(x)=x$, and $3\cdot (3\nvars^2-\delta(x))+1$ cliques otherwise. 

  Finally, if there is a clause $f_2(x)=x$ in the instance $\mathcal{I}$, then we delete the edge labelled by $x_{in}$ in the corresponding variable clique. 

	Observe that in the constructed instance of \quarkndel, among all the $9\nvars^2+2$ edges labelled by $x, x_{in}, x_{out}$, where $x$ is any variable, we have to delete either none, or all of them.
	This is because the deletion of any of them forces the deletion of all the others due to the appearance of induced copies of $K_n\setminus e$ in the graph.
	Moreover, if the edge $x_{in}$ is not present due to the existence of constraint $f_2(x)=x$ in $\mathcal{I}$, then all of them have to be deleted.

\begin{claim}\label{cl:knedel-there}
If there is a satisfying assignment with $k$ ones for the instance $\mathcal{I}$, then it is possible to delete $(9\nvars^2+2)\cdot k$ edges in $T(\mathcal{I})$ in order to make it a $K_n\setminus e$-free graph.
\end{claim}
\begin{proof}
It is enough to delete all edges labelled by $x, x_{in}, x_{out}$ for all variables $x$ that are set to $1$ in the satisfying assignment; the number of such edges is exactly $(9\nvars^2+2)\cdot k$. 
Let us prove the statement. Suppose the obtained graph is not $K_n\setminus e$-free. 
Let $H'$ be an induced subgraph isomorphic to $K_n\setminus e$. 
Note that for $n\geq 5$ the graph $K_n\setminus e$ is $3$-connected. 
Moreover, even after deletion of two arbitrary vertices in $K_n\setminus e$, there are no two vertices at distance larger than two. 
Consequently, a direct check shows that the assumed $H'$ subgraph must stay completely in one of the cliques corresponding to a constraint or to a variable, or 
in one of the cliques connecting a variable clique with a constraint clique.
Obviously, $H'$ cannot be contained in a variable clique or a connection clique, as in such cliques either all edges are present, or two edges are missing. 
This means that $H'$ must stay in a constraint clique, so exactly one of the edges of this constraint clique is deleted.
However, this is equivalent with the corresponding constraint being not satisfied under the considered assignment; this is a contradiction.
\cqed\end{proof}
	
\begin{claim}\label{cl:knedel-back}
If $T(\mathcal{I})$ admits a solution of cost $k'$, 
then there is a satisfying assignment for the instance~$\mathcal{I}$ with $\lfloor k' / (9\nvars^2+2) \rfloor$ ones.
\end{claim}
\begin{proof}
Take any solution for the output instance $T(\mathcal{I})$. 
As mentioned earlier, in any solution for $T(\mathcal{I})$, for any variable $x$ either all edges labeled by $x, x_{in}, x_{out}$ are deleted or none of them is deleted. 
The number of such edges for one variable $x$ is equal to $9\nvars^2+2$.
We set a variable to $1$ if and only if the corresponding edges are deleted in the considered solution for $T(\mathcal{I})$. 
All clauses of the form $f_2(x)$ will be satisfied, since in the construction of $T(\mathcal{I})$ 
 we delete $x_{in}$ if the clause $f_2(x)=x$ is present in $\mathcal{I}$.  
All $f_1$-type constraints will be satisfied as well, as otherwise in the clique corresponding to an unsatisfied constraint only one edge would be deleted and, hence, the graph would not be $K_n\setminus e$-free. 
\cqed\end{proof}

The correctness of the transformation follows from Claims~\ref{cl:knedel-there} and~\ref{cl:knedel-back}; hence the proof of Lemma~\ref{lem:horn-red1} is complete.
\end{proof}

\subsection{Lifting the quarantine}
\label{ssec:minhorn2}

In the following lemma we show how to reduce an instance of the quarantined
problem to its regular version, using the same
approach as in the proof of Lemma~\ref{thm:Quar-3con}.

\begin{lemma}\label{lem:horn-red2}
Let $n\geq 5$. There is a polynomial-time reduction
which, given an instance $G$ of \quarkndel
with $m$ edges, outputs an instance $G'$ of \kndel such that:
\begin{itemize}
\item $G'$ has $\Oh(m^3)$ vertices and edges.
\item If there is a solution of size $k$ for the instance $G$,
then there is a solution of size $k$ for the instance $G'$.
\item If there is a solution of size $k \leq m^2$ for the instance $G'$,
then there is a solution of size $k$ for the instance $G$.
\end{itemize}
\end{lemma}
\begin{proof}

We apply the reduction described in the proof of Lemma~\ref{thm:Quar-3con} for $p(m)=m^2$ and $H=K_n\setminus e$.
Now we verify that $G'$ has the claimed properties.
The bound on the size of $G'$ follows directly from the size bound given by Lemma~\ref{thm:Quar-3con}.

Suppose first that $G$ has some solution of size $k$. 
In the proof of Lemma~\ref{thm:Quar-3con} we argued that the same solution also works for the instance $G'$ (see the proof of Claim~\ref{cl:Hdel-quar-there}).
Hence, $G'$ also has a solution of size $k$.

Suppose now that $G'$ has a solution $F$ of some size $k\leq m^2$. 
In the proof of Claim~\ref{cl:Hdel-quar-back} we argued that $F$ does not delete any of the undeletable edges of $G$, because this would require deleting at least $m^2$ more edges in the attached gadgets.
Hence, $F\cap E(G)$ is a set of size at most $k$, whose deletion turns $G$ into an $H$-free graph, due to being an induced subgraph of $G'-F$.
Hence, $G$ has some solution of size at most $k$.
\end{proof}

%Note that the composition of the reductions of Lemmas~\ref{lem:horn-red1} and~\ref{lem:horn-red2}
%almost gives an $A$-reduction (for $\alpha=1$) from a \minhorn-hard problem \minones,
%yielding the hardness part of Theorem~\ref{thm:minhorn-new1}.
%The only technical caveat is that the last property from the 
%statement of Lemma~\ref{lem:horn-red2} holds only for $k \leq m^2$.
%Note that for large enough $m$, the composed reduction fulfills all the requirements
%for being an $A$-reduction assuming $k \le \sqrt{|E(G')|}$.
%Moreover, any solution for the instance of \quarkndel
%translates into a solution of \kndel with size smaller than $\sqrt{|E(G')|}$.
%Therefore, we decided to declare solutions for the final instance of \kndel
%infeasible whenever they are of size at least $\sqrt{|E(G')|}$,
%which is reflected by the exact statement of Theorem~\ref{thm:minhorn-new1}.

The composition of the reductions of Lemmas~\ref{lem:horn-red1} and~\ref{lem:horn-red2}
gives an $A$-reduction (for $\alpha=1$) from a \minhorn-hard problem \minones,
yielding the hardness part of Theorem~\ref{thm:minhorn-new1}.
Indeed, given an instance $\mathcal{I}$ of \minones we
can transform it into an instance $G$ of \quarkndel using Lemma~\ref{lem:horn-red1},
which in turn we can further transform into an instance $G'$ of \kndel using Lemma~\ref{lem:horn-red2}.
Given any feasible solution $F'$ for $G'$ 
we check whether $|F'| \le |E(G)|^2$. If this is the case,
we translate back the solution $F'$ into
a solution $F$ for $G$ (using Lemma~\ref{lem:horn-red2}) and then into a solution for the initial instance $\mathcal{I}$ (using Lemma~\ref{lem:horn-red1}).
On the other hand, if $|F'| > |E(G)|^2$, then we may take a trivial solution being an assignment setting all the variables to one.
This is an $r$-approximation where $r>|E(G)|$, as $|E(G)|>n_{vars}$ for the initial instance $\mathcal{I}$.  
The assignment will satisfy all the contraints and will be at least an $r$-approximation as we need to assign at least one variable to one, otherwise we may output all zeroes assignment. 

\subsection{Completeness}
\label{ssec:minhorn3}

To finish the proof of Theorem~\ref{thm:minhorn-new1} it remains to show a reduction in the other direction: from \kndel to \minhorn.
We achieve this goal by presenting an $A$-reduction from the \kndel problem
to another variant of \minones, which is \minhorn-complete.

\begin{definition} Let $n\geq 5$, and let $t=n(n-1)/2$. We define family of constraints $\mathcal{F}_n''=\{f_n, g_n\}$ as follows:
	\begin{itemize}
		\item $f_n(x_1, x_2, \dots x_t) = 0$ if and only if exactly one of the variables takes value 1;
		\item  $g_n(x_1, x_2, \dots x_{t-1})=0$ if and only if all the variables take value $0$.
	\end{itemize}
\end{definition}

The proof of the following lemma is a technical check that is essentially the same as the proof of Lemma~\ref{lem:fproperties}.
Hence, we leave it to the reader.

\begin{lemma} 
For each $n\geq 5$, the set of constraints $\mathcal{F}_n''=\{f_n, g_n\}$ is weakly positive, and at the same time it is neither $0$-valid, nor IHS-$B$ for any $B$.
\end{lemma}

Therefore, by Theorem~\ref{thm:cited} we know that \minonesn is \minhorn-complete
and it suffices to present an $A$-reduction
from \kndel to \minonesn.

\begin{lemma}\label{lem:del-to-horn}
There is a polynomial-time algorithm, which given
an instance $G$ of \kndel produces an instance $\mathcal{I}$ of \minonesn,
such that it is possible to remove exactly $k$ edges in $G$ to make it $K_n\setminus e$-free 
if and only if one can find a satisfying assignment for $\mathcal{I}$ that sets exactly $k$ variables to $1$.
\end{lemma}
\begin{proof}
Consider an instance $G$ of the \kndel problem. 
We enumerate all the edges in the graph $G$ as $e_1, e_2, \dots, e_m$, and to each edge $e_i$ we assign a fresh boolean variable $x_i$. 
For any induced subgraph $H$ isomorphic to $K_n\setminus e$ we list  all its edges $e_{i_1}, e_{i_2}, \dots, e_{i_{t-1}}$ and create a corresponding constraint $g(x_{i_1}, x_{i_2}, \dots, x_{i_{t-1}})$. 
For any induced clique $K$ containing $n$ vertices and edges $e_{i_1}, e_{i_2}, \dots, e_{i_t}$, we create a constraint $f(x_{i_1}, x_{i_2}, \dots, x_{i_t})$. 
The output instance $\mathcal{I}$ of \minonesn is obtained by taking $x_i$ to be the variable set, and putting all the constraints constructed above. 
	
Note that if we delete some edges in the graph $G$, then an induced copy of the graph $K_n\setminus e$ can be obtained only on vertices that originally were inducing  $K_n\setminus e$ or $K_n$. 
The constraints in the constructed instance guarantee that in each induced $K_n\setminus e$ subgraph at least one edge from the subgraph must be deleted, 
and in each induced subgraph $K_n$ either at least two edges should be deleted, or none of the edges should be deleted. 
So, for any $S\subseteq \{1,2,\ldots,|E(G)|\}$,
the graph $G-F$, where $F=\{e_i \colon i \in S \}$, is $K_n\setminus e$-free if and only if the assignment $\{x_i=1$ iff $i\in S\}$ satisfies $\mathcal{I}$.
This equivalence of solution sets immediately proves the lemma.
\end{proof}

As discussed earlier, Lemma~\ref{lem:del-to-horn} gives an $A$-reduction from \kndel to  \minonesn, which is \minhorn-complete, thereby proving that \kndel is $A$-reducible to \minhorn. 
This concludes the proof of Theorem~\ref{thm:minhorn-new1}.

%% file: kndelnew.pdf_tex
%% Creator: Inkscape 0.91_64bit, www.inkscape.org
%% Accompanies image file 'kndelnew.pdf' (pdf, eps, ps)
\begingroup%
  \makeatletter%
  \providecommand\color[2][]{%
    \errmessage{(Inkscape) Color is used for the text in Inkscape, but the package 'color.sty' is not loaded}%
    \renewcommand\color[2][]{}%
  }%
  \providecommand\transparent[1]{%
    \errmessage{(Inkscape) Transparency is used (non-zero) for the text in Inkscape, but the package 'transparent.sty' is not loaded}%
    \renewcommand\transparent[1]{}%
  }%
  \providecommand\rotatebox[2]{#2}%
  \ifx\svgwidth\undefined%
    \setlength{\unitlength}{640bp}%
    \ifx\svgscale\undefined%
      \relax%
    \else%
      \setlength{\unitlength}{\unitlength * \real{\svgscale}}%
    \fi%
  \else%
    \setlength{\unitlength}{\svgwidth}%
  \fi%
  \global\let\svgwidth\undefined%
  \global\let\svgscale\undefined%
  \makeatother%
  \begin{picture}(1,0.75)%
    \put(0,0){\includegraphics[width=\unitlength,page=1]{kndelnew.pdf}}%
    \put(0.2162479,0.07142555){\color[rgb]{0,0,0}\makebox(0,0)[lb]{\smash{ $x_{in}$}}}%
    \put(0.2912479,0.25892555){\color[rgb]{0,0,0}\makebox(0,0)[lb]{\smash{$x_{out}$}}}%
    \put(0.4162479,0.33392555){\color[rgb]{0,0,0}\makebox(0,0)[lb]{\smash{$x$}}}%
    \put(0.5287479,0.40892555){\color[rgb]{0,0,0}\makebox(0,0)[lb]{\smash{$x$}}}%
    \put(0.66624786,0.50892557){\color[rgb]{0,0,0}\makebox(0,0)[lb]{\smash{$x$}}}%
    \put(0.72874786,0.43392555){\color[rgb]{0,0,0}\makebox(0,0)[lb]{\smash{$y$}}}%
    \put(0.82874786,0.49642557){\color[rgb]{0,0,0}\makebox(0,0)[lb]{\smash{$z$}}}%
    \put(0.3375,0.1625){\color[rgb]{0,0,0}\makebox(0,0)[lb]{\smash{clique for}}}%
    \put(0.3375,0.1375){\color[rgb]{0,0,0}\makebox(0,0)[lb]{\smash{}}}%
    \put(0.3190641,0.13160744){\color[rgb]{0,0,0}\makebox(0,0)[lb]{\smash{variable $x$}}}%
    \put(0.68678513,0.7){\color[rgb]{0,0,0}\makebox(0,0)[lb]{\smash{clique for}}}%
    \put(0.64379547,0.66616117){\color[rgb]{0,0,0}\makebox(0,0)[lb]{\smash{constraint $f_1(x,y,z)$}}}%
    \put(0.2125,0.4125){\color[rgb]{0,0,0}\makebox(0,0)[lb]{\smash{first clique}}}%
    \put(0.525,0.25){\color[rgb]{0,0,0}\makebox(0,0)[lb]{\smash{second clique}}}%
    \put(0.4375,0.6){\color[rgb]{0,0,0}\makebox(0,0)[lb]{\smash{third clique}}}%
  \end{picture}%
\endgroup%

%% file: specific.tex
\section{Specific constructions for short paths and cycles}\label{sec:specific}

In this section we extend the general results yielded by Theorems~\ref{thm:Quar-3con} and~\ref{thm:Quar-3con-compl} in the direction of obtaining a full picture of the approximation complexity 
for $H$ being a path or a cycle.
It can be easily seen that the complements of $P_\ell$ and $C_\ell$ for $\ell\geq 6$ satisfy the preconditions of Theorems~\ref{thm:Quar-3con} and~\ref{thm:Quar-3con-compl}. 
Hence, by complementation, we have already established hardness of approximation for these cases.
We are left with considering the edge modification problems for $H=P_\ell$ and $H=C_\ell$ for $\ell\leq 5$.
Therefore, to complete the proof of Theorem~\ref{thm:main-paths-cycles}, it remains to prove the following.

\begin{lemma}\label{lem:specific}
Let $H$ be equal to $C_4$, $C_5$, or $P_5$. 
Then neither \Hfreedel nor \Hfreecom admits a $\poly(\OPT)$-factor approximation algorithm working in polynomial time, unless $\mathrm{P}=\mathrm{NP}$.
Moreover, unless ETH fails, there is even no $\poly(\OPT)$-factor approximation algorithm working in time $2^{o(\OPT)}\cdot n^{\Oh(1)}$, for any of these problems.
\end{lemma}

Before we proceed to the proof of the missing cases (Lemma~\ref{lem:specific}), let us check that we indeed 
obtain a full classification for cycles, and an almost full classification for paths, as promised in the introduction.
The problem {\sc{$C_3$-Free Edge Deletion}}, aka {\sc{Triangle-Free Edge Deletion}}, admits a trivial greedy $3$-approximation algorithm, whereas
{\sc{$P_3$-Free Edge Deletion}}, aka {\sc{Cluster Edge Deletion}}, admits a constant-factor approximation algorithm given by Natanzon~\cite{natanzonmaster}.
The problem {\sc{$C_3$-Free Edge Completion}} has no sense, and {\sc{$P_3$-Free Edge Completion}} is polynomial-time solvable because there is only one way to destroy every obstacle.
The only missing case is {\sc{$P_4$-Free Edge Deletion}}, which is equivalent to {\sc{$P_4$-Free Edge Completion}} by complementation.

\begin{figure}
    \centering
        \def\svgwidth{0.3\textwidth}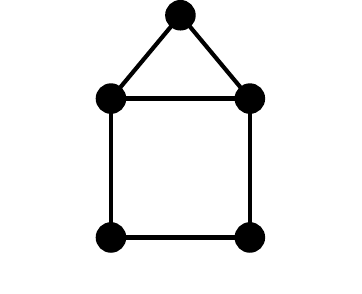
    \caption{House: the complement of a $P_5$.}\label{fig:house}
\end{figure}

The rest of this section is devoted to the proof of Lemma~\ref{lem:specific}. 
For this, we implement the same strategy as in Theorems~\ref{thm:Quar-3con} and~\ref{thm:Quar-3con-compl}: 
we first prove hardness of sandwich problems by giving linear reductions from {\sc{3SAT}}, and then we reduce to the standard optimization variant by introducing the approximation gap.
For convenience, instead of working with {\sc{$P_5$-Free Edge Deletion}} and {\sc{$P_5$-Free Edge Completion}}, we respectively consider {\sc{House-Free Edge Completion}} and {\sc{House-Free Edge Deletion}},
where house is the complement of $P_5$: a $4$-cycle with a triangle built on one of the edges (see Figure~\ref{fig:house}). 
These problems are equivalent to the ones concerning $P_5$-s by complementation of the instance.
Also, observe that {\sc{$C_5$-Free Edge Deletion}} and {\sc{$C_5$-Free Edge Completion}} are equivalent by complementation, and hence we consider only the former.

\subsection{Sandwich deletion problems}

We start with the hardness proof for {\sc{Sandwich $C_4$-Free Edge Deletion}}, which will serve as a template for other reductions. 
The structural property of the instance, described in the statement, will turn out to be useful in some further arguments.

\begin{lemma}\label{lem:qC4-del}
There is a polynomial-time reduction which, given an instance of {\sc{3SAT}} with $n$ variables and $m$ clauses, 
constructs an equivalent instance $G$ of {\sc{Sandwich $C_4$-Free Edge Deletion}} with $\Oh(n+m)$ vertices and edges.
Moreover, $G$ has the following additional property: every (not necessarily induced) $C_4$ subgraph of $G$ contains an undeletable edge.
Consequently, {\sc{Sandwich $C_4$-Free Edge Deletion}} is $\mathrm{NP}$-hard, even on such instances.
\end{lemma}
\begin{proof}
Let $\varphi$ be the given formula in 3CNF, and let $\vars$ and $\cls$ be the sets of variables and clauses of $\varphi$. 
By standard modifications of the formula, we may assume that each clause contains exactly three literals of pairwise different variables.

\begin{figure}
    \centering
    \subfloat[Variable gadget $G^\vars$]{
        \def\svgwidth{0.4\textwidth}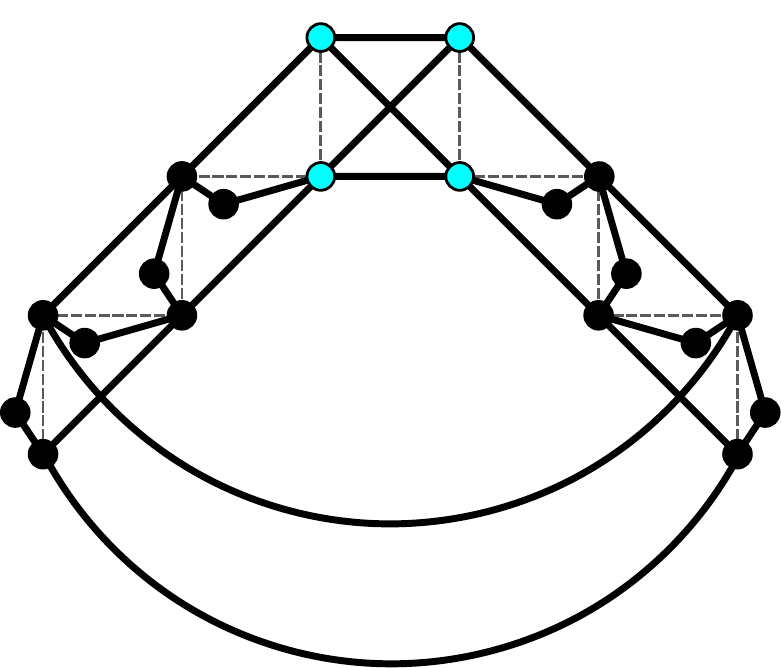
    }
    \quad
    \subfloat[Clause gadget $H^\cls$]{
	\raisebox{0.8cm}{\def\svgwidth{0.26\textwidth}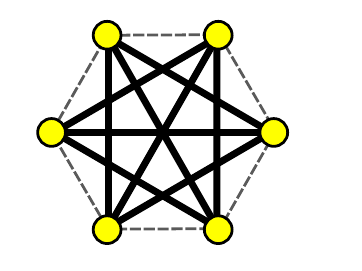}
    }
    \quad
    \subfloat[Connector gadget]{
	\raisebox{0.9cm}{\def\svgwidth{0.23\textwidth}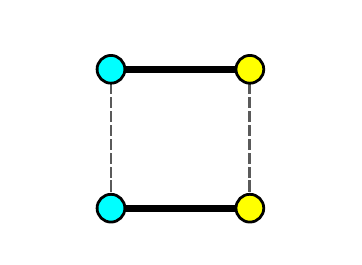}
    }
    \caption{Gadgets for {\sc{Sandwich $C_4$-Free Edge Deletion}}.}\label{fig:c4-del-gadgets}
\end{figure}

We introduce gadgets for variables, for clauses, and for connections between variable and clause gadgets. 
They are depicted in Figure~\ref{fig:c4-del-gadgets}, where thick edges are undeletable and dashed edges are deletable.
The variable gadget $G^\vars$, depicted on the first panel, has four named vertices $u_\top$, $v_\top$, $u_\bot$ and $v_\bot$, which will be used to connect the copies of this gadget to the rest of the construction.
The properties of the variable gadget are described in the following claim. Its proof follows by a direct check, and hence is omitted.

\begin{claim}\label{cl:c4-del-var}
There are exactly two solutions to the {\sc{Sandwich $C_4$-Free Edge Deletion}} instance $G^\vars$. One of them, denoted $F_\top$, contains $u_\top v_\top$ and does not contain $u_\bot v_\bot$, and the second, denoted
$F_\bot$, contains $u_\bot v_\bot$ and does not contain $u_\top v_\top$.
\end{claim}

Next, we describe the clause gadget $H^\cls$, depicted on the second panel of Figure~\ref{fig:c4-del-gadgets}. 
It consists of a clique on $6$ vertices $\{s_1,t_1,s_2,t_2,s_3,t_3\}$, where the cycle $s_1-t_1-s_2-t_2-s_3-t_3-s_1$ has deletable edges, and all the other edges are undeletable. 
Again, the properties of the clause gadget are described in the following claim, whose proof is omitted due to being straightforward.

\begin{claim}\label{cl:c4-del-cls}
In the {\sc{Sandwich $C_4$-Free Edge Deletion}} instance $H^\cls$ there is no solution that simultaneously contains all three edges $s_1t_1$, $s_2t_2$ and $s_3t_3$. 
However, for each $i=1,2,3$, there is a solution $F_i$ that does not contain $s_it_i$, but contains both the other edges from this triple.
\end{claim}

For every variable $x\in \vars$ we create a copy $G^x$ of the variable gadget $G^\vars$. 
The copies of vertices $u_\top$, $v_\top$, $u_\bot$ and $v_\bot$ in $G^x$ are respectively renamed to $u^x_\top$, $v^x_\top$, $u^x_\bot$ and $v^x_\bot$.
For every clause $c\in \cls$ we create a copy $H^c$ of the clause gadget $H^\cls$.
The copies of vertices $s_1,t_1,s_2,t_2,s_3,t_3$ in $H^c$ are respectively renamed to $s^c_1,t^c_1,s^c_2,t^c_2,s^c_3,t^c_3$.

Finally, we wire the variable gadgets and clause gadgets using connector gadgets, which are just $C_4$-s (depicted on the third panel of Figure~\ref{fig:c4-del-gadgets}).
More precisely, whenever $x$ appears in the $i$-th literal clause $c$, we connect $s_i^c$ with $u_p^x$ and $t_i^c$ with $v_p^x$ using undeletable edges, 
where $p=\top$ if the appearance of $x$ in $c$ is positive, and $p=\bot$ if it is negative.
Note that the deletable edges $uv$ and $st$ depicted in Figure~\ref{fig:c4-del-gadgets} are always present in respective variable or clause gadgets

This concludes the construction; the constructed graph will be denoted by $G$. Obviously $G$ has $\Oh(n+m)$ vertices and edges.
It is straightforward to see that the asserted structural property of $G$ is satisfied: the subgraph spanned by deletable edges consists of disjoint paths and cycles on $6$ vertices, 
hence every $C_4$ subgraph must contain at least one undeletable edge.

We now need to verify that the obtained instance $G$ of {\sc{Sandwich $C_4$-Deletion}} has a solution if and only if the input formula $\varphi$ is satisfiable.
For this, the following claim will be useful; its proof is a straightforward check following from the fact
that each vertex of a clause gadget is incident with at most one edge leading to a variable gadget, and hence we omit the proof.

\begin{claim}\label{cl:c4-only-within}
Every (not necessarily induced) $C_4$ in $G$ is entirely contained in one variable gadget, in one clause gadget, or forms one connector gadget.
\end{claim}

Suppose first that $\alpha\colon \vars \to \{\bot,\top\}$ is a variable assignment that satisfies $\varphi$.
Construct a subset $F$ of deletable edges in $G$ as follows:
\begin{itemize}
\item For each variable $x\in \vars$, add to $F$ the solution $F_{\alpha(x)}$ in the variable gadget $G^x$, given by Claim~\ref{cl:c4-del-var}.
\item For each clause $c\in \cls$, arbitrarily choose an index $i_c\in \{1,2,3\}$ of any of its literal that satisfies it under $\alpha$; such literal exists due to $\alpha$ being a satisfying assignment.
Then add to $F$ the solution $F_{i_c}$ in the clause gadget $H^c$, given by Claim~\ref{cl:c4-del-cls}.
\end{itemize}
By Claim~\ref{cl:c4-only-within}, to verify the $G-F$ is $C_4$-free, it suffices to show that there is no induced $C_4$ within any variable gadget or within any clause gadget, and that one of the edges in each
connector gadget is removed. The first two checks follow immediately from Claims~\ref{cl:c4-del-var} and~\ref{cl:c4-del-cls}. For the last check, fix some clause $c$ and variable $x$ appearing in it;
we examine the connector gadget between $G^x$ and $H^c$.
Suppose that $x$ appears in the $i$-th literal of $c$, and assume w.l.o.g. that this appearance is positive; the second case is symmetric. 
If $\alpha(x)=\top$, then the edge $u^x_\top v^x_\top$ is deleted in $G^x$, and hence the $C_4$
in the connector gadget is destroyed. Otherwise $\alpha(x)=\bot$, and hence the literal containing $x$ cannot satisfy the clause $c$ under assignment $\alpha$. 
From the construction of $F$ it follows that the edge $s_i^ct_i^c$ is deleted in the gadget $H^c$, and hence the $C_4$ in the connector gadget is also destroyed.

For the other direction, suppose that there is a subset $F$ of deletable edges in $G$ such that $G-F$ is $C_4$-free.
By Claim~\ref{cl:c4-del-var}, the intersection of $F$ with the edge set of each variable gadget $G^x$ must be equal either to solution $F_\top$ or to solution $F_\bot$.
Define assignment $\alpha\colon \vars \to \{\bot,\top\}$ as follows: $\alpha(x)=\top$ if this intersection is $F_\top$, and $\alpha(x)=\bot$ if it is $F_\bot$.
In particular, edge $u^x_\top v^x_\top$ belongs to $F$ if and only if $\alpha(x)=\top$, and the symmetric claim holds also for $u^x_\bot v^x_\bot$.
We verify that $\alpha$ is a satisfying assignment for $\varphi$.
Take any clause $c\in \cls$, and for the sake of contradiction suppose it is not satisfied under $\alpha$.
By the construction of $\alpha$, this means that in all three connector gadgets connecting $H^c$ with variable gadgets of variables appearing in $c$, the deletable edges from the variable gadgets are not included in $F$.
Since each connector gadget induces a $C_4$ with only two edges deletable, it follows that all three edges $s^c_1t^c_1$, $s^c_2t^c_2$, and $s^c_3t^c_3$ have to be included in $F$.
However, Claim~\ref{cl:c4-del-cls} asserts that there is no solution within the clause gadget $H^c$ that simultaneously contains all these three edges.
This is a contradiction, and hence we conclude that assignment $\alpha$ satisfies formula~$\varphi$.
\end{proof}

We now move to the proof for {\sc{Sandwich $C_5$-Free Edge Deletion}}, which is a minor modification of the construction for {\sc{Sandwich $C_4$-Free Edge Deletion}}. 
For this reason, we only sketch how the construction need to be modified, and argue that the correctness proof follows the same steps.

\begin{lemma}\label{lem:qC5-del}
There is a polynomial-time reduction which, given an instance of {\sc{3SAT}} with $n$ variables and $m$ clauses, 
constructs an equivalent instance $G$ of {\sc{Sandwich $C_5$-Free Edge Deletion}} with $\Oh(n+m)$ edges.
Consequently, {\sc{Sandwich $C_5$-Free Edge Deletion}} is $\mathrm{NP}$-hard.
\end{lemma}
\begin{proof}
We perform essentially the same construction as in the proof of Lemma~\ref{lem:qC4-del}, 
but we replace the variable, clause and connector gadgets with $C_5$-specific constructions depicted in Figure~\ref{fig:c5-del-gadgets}.

\begin{figure}
    \centering
    \subfloat[Variable gadget $G_\vars$]{
        \def\svgwidth{0.4\textwidth}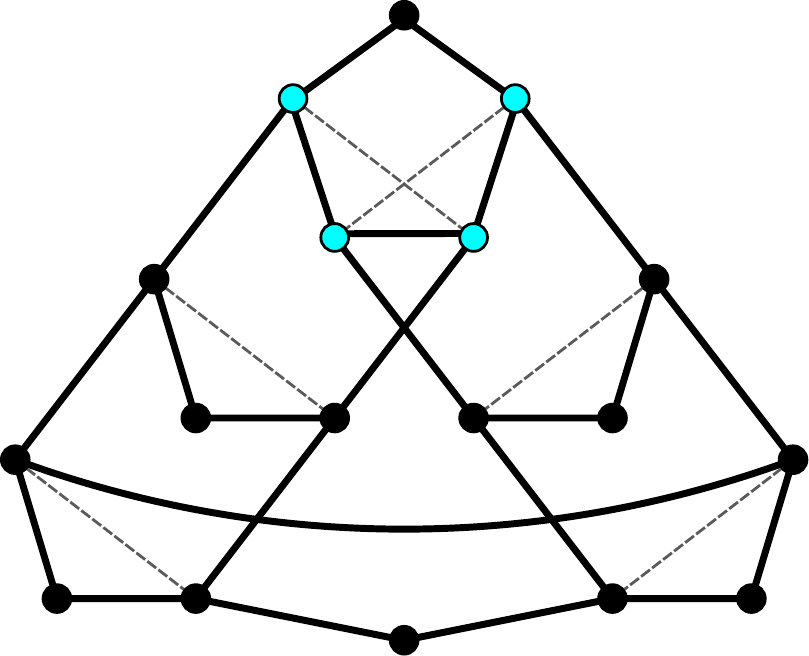
    }
    \quad
    \subfloat[Clause gadget $H_\cls$]{
	\raisebox{0.8cm}{\def\svgwidth{0.26\textwidth}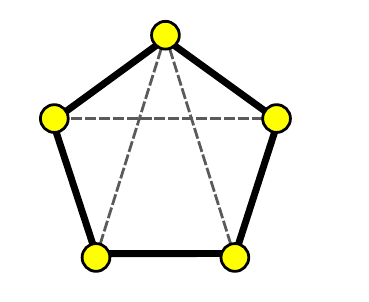}
    }
    \quad
    \subfloat[Connector gadget]{
	\raisebox{0.9cm}{\def\svgwidth{0.23\textwidth}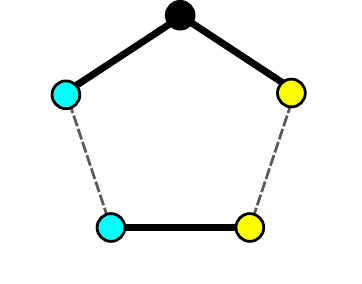}
    }
    \caption{Gadgets for {\sc{Sandwich $C_5$-Deletion}}.}\label{fig:c5-del-gadgets}
\end{figure}

The variable gadget $G^\vars$ is depicted on the first panel of Figure~\ref{fig:c5-del-gadgets}. As before, it has four named vertices: $u_\top$, $v_\top$, $u_\bot$, and $v_\bot$.
Again, a direct check, whose proof is omitted, yields the following.

\begin{claim}\label{cl:c5-del-var}
There are exactly two solutions to the {\sc{Sandwich $C_5$-Free Edge Deletion}} instance $G^\vars$. One of them, denoted $F_\top$, contains $u_\top v_\top$ and does not contain $u_\bot v_\bot$, and the second, denoted
$F_\bot$, contains $u_\bot v_\bot$ and does not contain $u_\top v_\top$.
\end{claim}

The clause gadget $H^\cls$ is depicted on the second panel of Figure~\ref{fig:c5-del-gadgets}. 
It has five vertices, but in order to keep the description same as in Lemma~\ref{lem:qC4-del}, one of them is named both $s_1$ and $s_2$.
Thus, the gadget has three deletable edges $s_1t_1$, $s_2t_2$, and $s_3t_3$.
Again, a direct check, whose proof is omitted, yields the following.

\begin{claim}\label{cl:c5-del-cls}
In the {\sc{Sandwich $C_5$-Free Edge Deletion}} instance $H^\cls$ there is no solution that simultaneously contains all three edges $s_1t_1$, $s_2t_2$ and $s_3t_3$. 
However, for each $i=1,2,3$, there is a solution $F_i$ that does not contain $s_it_i$, but contains both the other edges from this triple.
\end{claim}

As in the proof of Lemma~\ref{lem:qC4-del}, we create one variable gadget $G^x$ for each variable $x$, and one clause gadget $H^c$ for each clause $c$.
We follow the same renaming convention, where the variable/clause corresponding to the gadget is in the superscript of each vertex of this gadget.
The variable and clause gadgets are connected to each other via connector gadgets exactly as in Lemma~\ref{lem:qC4-del}, 
which this time are simply $C_5$-s (see the third panel of Figure~\ref{fig:c5-del-gadgets}): 
the appropriate vertex $s$ is connected to the appropriate vertex $u$ via a path of length $2$, and the appropriate vertex $t$ is connected to the appropriate vertex $v$ via a single edge; 
all these edges are undeletable. Similarly as in the proof of Lemma~\ref{lem:qC4-del}, a direct check yields the following.

\begin{claim}\label{cl:c5-only-within}
Every (not necessarily induced) $C_5$ in $G$ is entirely contained in one variable gadget, in one clause gadget, or forms one connector gadget.
\end{claim}

We remark that for the check of Claim~\ref{cl:c5-only-within} it is important that the vertex $s_1=s_2$ in the clause gadget that is shared between two deletable edges, 
is always the endpoint of the path of length $2$, not $1$, in the corresponding connector gadgets connecting it to variable gadgets.

Having Claims~\ref{cl:c5-del-var},~\ref{cl:c5-del-cls}, and~\ref{cl:c5-only-within} in place, the proof of the correctness is exactly the same as in the proof of Lemma~\ref{lem:qC5-del}.
We leave the easy verification to the reader.
\end{proof}

For now, we postpone the argumentation for the remaining deletion problem, namely {\sc{House-Free Edge Deletion}}. We will deal with this case later, using a different reasoning.

\subsection{Sandwich completion problems}

We now proceed with proving the hardness of sandwich variants of the relevant completion problems: {\sc{$C_4$-Free Edge Completion}} and {\sc{House-Free Edge Completion}}.

\begin{lemma}\label{lem:qC4-comp}
There is a polynomial-time reduction which, given an instance of {\sc{3SAT}} with $n$ variables and $m$ clauses, 
constructs an equivalent instance $G$ of {\sc{Sandwich $C_4$-Free Edge Completion}} with $\Oh(n+m)$ vertices, edges, and fillable non-edges. 
Moreover, $G$ has the following additional property: the graph spanned by fillable non-edges does not contain any (not necessarily induced) $C_4$.
Consequently, {\sc{Sandwich $C_4$-Free Edge Completion}} is $\mathrm{NP}$-hard, even on such instances.
\end{lemma}
\begin{proof}
We modify slightly the reduction of Drange et al.~\cite{DrangeFPV15}, which shows that (the minimization variant of) {\sc{$C_4$-Free Edge Completion}} has no subexponential-time algorithm, under the assumption of ETH.
Unfortunately, while this construction happens to basically work ``as is'' in our setting, the proof of its correctness, contained in~\cite{DrangeFPV15}, uses budget constraints for convenience.
For this reason, we now recall the whole construction, perform slight modifications to adjust it to the sandwich setting, and argue its correctness.

Let $\varphi$ be the given formula in 3CNF, and let $\vars$ and $\cls$ be the sets of variables and clauses of $\varphi$. 
By standard modifications of the formula we may assume that each clause contains exactly three literals of pairwise different variables.
For a variable $x$, by $p_x$ we denote the number of occurrences of $x$ in $\varphi$. 
By copying the whole formula several times, we may assume that $p_x\geq 2$ for each $x\in \vars$.

For each variable $x$, we construct a variable gadget $G^x$ depicted in Figure~\ref{fig:c4compl-variable-gadget-enlarged}; this gadget is exactly the same as in~\cite{DrangeFPV15}, 
and in particular the figures depicting it are taken verbatim from~\cite{DrangeFPV15} by the consent of the authors.
The gadget consists of two cycles of length $4p_x$: 
$$
t^x_0 - t^x_1 - \ldots -t^x_{p_x-1} - t^x_0\qquad  \textrm{and}\qquad b^x_0 - b^x_1 - \ldots -b^x_{p_x-1} - b^x_0,
$$
connected into a cyclic ``ladder'' by adding edges $t^x_ib^x_i$, for all $i=0,1,\ldots,p_x-1$.
Moreover, for each $i=0,1,\ldots,p_x-1$ we introduce vertices $u^x_i$ and $d^x_i$. 
We make $u^x_i$ adjacent to $t^x_{i-1}$, $t^x_{i}$, and $t^x_{i+1}$ (the indices behave cyclically modulo $4p_x$), whereas $b^x_i$ is made adjacent to $d^x_{i-1}$, $d^x_{i}$, and $d^x_{i+1}$.
In the constructed sandwich instance, within the gadget $G_x$ we declare only the diagonals of the $C_4$-s to be fillable, i.e., edges $t^x_{i}b^x_{i+1}$ and $t^x_{i+1}b^x{i}$ for $i=0,1,\ldots,p_x-1$.
All the other non-edges cannot be filled.

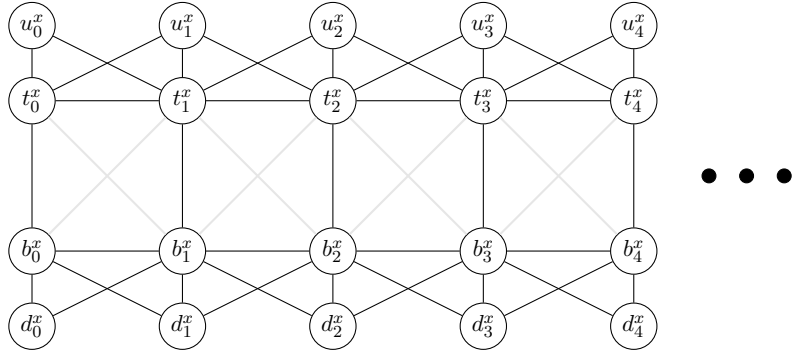
\begin{figure}[t]
  \centering
  \begin{tikzpicture}[every
    node/.style={circle,draw,scale=.8},scale=1,minimum size=2em, inner
    sep=1pt]
    \node (u0) at (-1,4) {$u^x_{0}$};
    \node (t0) at (-1,3) {$t^x_{0}$};
    \node (b0) at (-1,1) {$b^x_{0}$};
    \node (d0) at (-1,0) {$d^x_{0}$};

    \node (u1) at (1,4) {$u^x_{1}$};
    \node (t1) at (1,3) {$t^x_{1}$};
    \node (b1) at (1,1) {$b^x_{1}$};
    \node (d1) at (1,0) {$d^x_{1}$};

    \node (u2) at (3,4) {$u^x_{2}$};
    \node (t2) at (3,3) {$t^x_{2}$};
    \node (b2) at (3,1) {$b^x_{2}$};
    \node (d2) at (3,0) {$d^x_{2}$};

    \node (u3) at (5,4) {$u^x_{3}$};
    \node (t3) at (5,3) {$t^x_{3}$};
    \node (b3) at (5,1) {$b^x_{3}$};
    \node (d3) at (5,0) {$d^x_{3}$};

    \node (u4) at (7,4) {$u^x_{4}$};
    \node (t4) at (7,3) {$t^x_{4}$};
    \node (b4) at (7,1) {$b^x_{4}$};
    \node (d4) at (7,0) {$d^x_{4}$};

    \node[scale=.3,fill] (dot1) at (8.0, 2) {};
    \node[scale=.3,fill] (dot2) at (8.5, 2) {};
    \node[scale=.3,fill] (dot3) at (9.0, 2) {};

    \draw (u0) -- (t0) -- (b0) -- (d0);
    \draw (u1) -- (t1) -- (b1) -- (d1);
    \draw (u2) -- (t2) -- (b2) -- (d2);
    \draw (u3) -- (t3) -- (b3) -- (d3);
    \draw (u4) -- (t4) -- (b4) -- (d4);
    \draw (t0) -- (t1) -- (t2) -- (t3) -- (t4);
    \draw (b0) -- (b1) -- (b2) -- (b3) -- (b4);
    
    \draw (t0) -- (u1) -- (t2) -- (u3) -- (t4);
    \draw (u0) -- (t1) -- (u2) -- (t3) -- (u4);

    \draw (b0) -- (d1) -- (b2) -- (d3) -- (b4);
    \draw (d0) -- (b1) -- (d2) -- (b3) -- (d4);
    
    \draw[color=gray!20, thick] (t0) -- (b1);
    \draw[color=gray!20, thick] (b0) -- (t1);
    \draw[color=gray!20, thick] (t1) -- (b2);
    \draw[color=gray!20, thick] (b1) -- (t2);
    \draw[color=gray!20, thick] (t2) -- (b3);
    \draw[color=gray!20, thick] (b2) -- (t3);
    \draw[color=gray!20, thick] (t3) -- (b4);
    \draw[color=gray!20, thick] (b3) -- (t4);
    
  \end{tikzpicture}
  \caption{Variable gadget $G^x$; the light grey lines represent fillable edges. The figure is taken almost verbatim from Drange et al.~\cite{DrangeFPV15}, by the consent of the authors.}
  \label{fig:c4compl-variable-gadget-enlarged}
\end{figure}

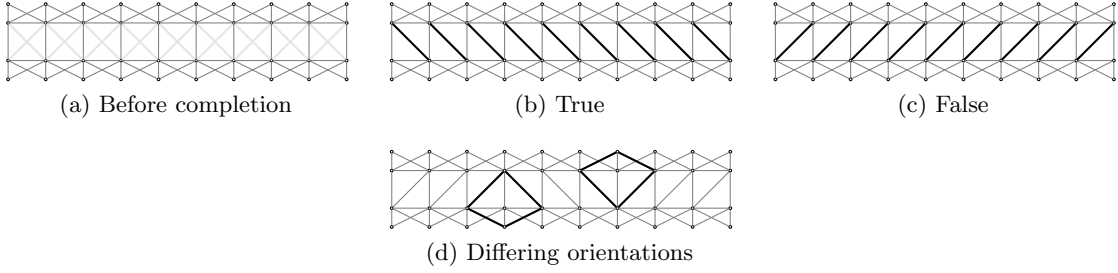
\begin{figure}[t]
  \centering
  \subfloat[Before completion] {
    \centering
    \begin{tikzpicture}[every
      node/.style={circle,draw,scale=.1},scale=.5]
      \foreach \s in {1,...,10}
      {
        \node (\s+u) at (\s,2) {};
        \node (\s+t) at (\s,1.5) {};
        \node (\s+b) at (\s,.5) {};
        \node (\s+d) at (\s,0) {};
        \draw[color=gray] (\s+u) -- (\s+t);
        \draw[color=gray] (\s+t) -- (\s+b);
        \draw[color=gray] (\s+b) -- (\s+d);
      }
      \foreach \s/\t in {1/2, 2/3, 3/4, 4/5, 5/6, 6/7, 7/8, 8/9, 9/10}
      {
        \draw[color=gray] (\s+u) -- (\t+t);
        \draw[color=gray] (\s+t) -- (\t+u);
        \draw[color=gray] (\s+t) -- (\t+t);
        \draw[color=gray] (\s+b) -- (\t+b);
        \draw[color=gray] (\s+b) -- (\t+d);
        \draw[color=gray] (\s+d) -- (\t+b);
        
        \draw[color=gray!20, thick] (\s+t) -- (\t+b);
        \draw[color=gray!20, thick] (\s+b) -- (\t+t);
      }
    \end{tikzpicture}
    \label{fig:c4compl-variable-gadget}
  }
  \hspace{.01\textwidth}
  \subfloat[True] {
    \centering
    \begin{tikzpicture}[every node/.style={circle,draw,scale=.1},scale=.5]
      \foreach \s in {1,...,10}
      {
        \node (\s+u) at (\s,2) {};
        \node (\s+t) at (\s,1.5) {};
        \node (\s+b) at (\s,.5) {};
        \node (\s+d) at (\s,0) {};
        \draw[color=gray] (\s+u) -- (\s+t);
        \draw[color=gray] (\s+t) -- (\s+b);
        \draw[color=gray] (\s+b) -- (\s+d);
      }
      \foreach \s/\t in {1/2, 2/3, 3/4, 4/5, 5/6, 6/7, 7/8, 8/9, 9/10}
      {
        \draw[color=gray] (\s+u) -- (\t+t);
        \draw[color=gray] (\s+t) -- (\t+u);
        \draw[color=gray] (\s+t) -- (\t+t);
        \draw[color=gray] (\s+b) -- (\t+b);
        \draw[color=gray] (\s+b) -- (\t+d);
        \draw[color=gray] (\s+d) -- (\t+b);
        \draw[thick] (\s+t) -- (\t+b); % TRUE
      }
    \end{tikzpicture}
  }
  \hspace{.01\textwidth}
  \subfloat[False]{
    \centering
    \begin{tikzpicture}[every node/.style={circle,draw,scale=.1},scale=.5]
      \foreach \s in {1,...,10}
      {
        \node (\s+u) at (\s,2) {};
        \node (\s+t) at (\s,1.5) {};
        \node (\s+b) at (\s,.5) {};
        \node (\s+d) at (\s,0) {};
        \draw[color=gray] (\s+u) -- (\s+t);
        \draw[color=gray] (\s+t) -- (\s+b);
        \draw[color=gray] (\s+b) -- (\s+d);
      }
      \foreach \s/\t in {1/2, 2/3, 3/4, 4/5, 5/6, 6/7, 7/8, 8/9, 9/10}
      {
        \draw[color=gray] (\s+u) -- (\t+t);
        \draw[color=gray] (\s+t) -- (\t+u);
        \draw[color=gray] (\s+t) -- (\t+t);
        \draw[color=gray] (\s+b) -- (\t+b);
        \draw[color=gray] (\s+b) -- (\t+d);
        \draw[color=gray] (\s+d) -- (\t+b);
        \draw[thick] (\s+b) -- (\t+t); % FALSE
      }
    \end{tikzpicture}
  }
  \hspace{.01\textwidth}
  \subfloat[Differing orientations]{
    \centering
    \begin{tikzpicture}[every node/.style={circle,draw,scale=.1},scale=.5]
      \foreach \s in {1,...,10}
      {
        \node (\s+u) at (\s,2) {};
        \node (\s+t) at (\s,1.5) {};
        \node (\s+b) at (\s,.5) {};
        \node (\s+d) at (\s,0) {};
        \draw[color=gray] (\s+u) -- (\s+t);
        \draw[color=gray] (\s+t) -- (\s+b);
        \draw[color=gray] (\s+b) -- (\s+d);
      }
      \foreach \s/\t in {1/2, 2/3, 3/4}
      {
        \draw[color=gray] (\s+u) -- (\t+t);
        \draw[color=gray] (\s+t) -- (\t+u);
        \draw[color=gray] (\s+t) -- (\t+t);
        \draw[color=gray] (\s+b) -- (\t+b);
        \draw[color=gray] (\s+b) -- (\t+d);
        \draw[color=gray] (\s+d) -- (\t+b);
        \draw[color=gray] (\s+b) -- (\t+t); % FALSE
      }
      \foreach \s/\t in {4/5, 5/6, 6/7}
      {
        \draw[color=gray] (\s+u) -- (\t+t);
        \draw[color=gray] (\s+t) -- (\t+u);
        \draw[color=gray] (\s+t) -- (\t+t);
        \draw[color=gray] (\s+b) -- (\t+b);
        \draw[color=gray] (\s+b) -- (\t+d);
        \draw[color=gray] (\s+d) -- (\t+b);
        \draw[color=gray] (\s+t) -- (\t+b); % TRUE
      }\foreach \s/\t in {7/8, 8/9, 9/10}
      {
        \draw[color=gray] (\s+u) -- (\t+t);
        \draw[color=gray] (\s+t) -- (\t+u);
        \draw[color=gray] (\s+t) -- (\t+t);
        \draw[color=gray] (\s+b) -- (\t+b);
        \draw[color=gray] (\s+b) -- (\t+d);
        \draw[color=gray] (\s+d) -- (\t+b);
        \draw[color=gray] (\s+b) -- (\t+t); % FALSE
      }
      % FALSE -- TRUE
      \draw[thick] (4+t) -- (5+b) -- (4+d) -- (3+b) -- (4+t);
      % TRUE -- FALSE
      \draw[thick] (7+b) -- (8+t) -- (7+u) -- (6+t) -- (7+b);
    \end{tikzpicture}
    \label{fig:c4compl-variable-gadget-differing}
  }
  \caption{The variable gadget $G^x$: before the completion, and after the completions corresponding to setting the variable to true/false.
  The light grey lines represent fillable non-edges.
  The last panel shows how differing orientations of the completed diagonals lead to $C_4$s that cannot be destroyed.
  The figure is taken almost verbatim from Drange et al.~\cite{DrangeFPV15}, by the consent of the authors.}
  \label{fig:c4compl-variable-gadgets}
\end{figure}

The following claim verifies that the constructed gadget has exactly two solutions. We remark that the proof of Drange et al.~\cite{DrangeFPV15} used at this point the budget constraints.

\begin{claim}\label{cl:c4-compl-var}
The {\sc{Sandwich $C_4$-Free Edge Completion}} instance $G^x$ has exactly two solutions, depicted on the second and third panel of Figure~\ref{fig:c4compl-variable-gadgets}.
The solution that takes all edges of the form $t^x_{i}b^x_{i+1}$ will be denoted by $F^x_\top$, whereas the solution that takes all edges of the form $t^x_{i+1}b^x_{i}$ will be denoted by $F^x_\bot$.
\end{claim}
\begin{proof}
For $i=0,1,\ldots,p_x-1$, let $W_i=\{b^x_i,b^x_{i+1},t^x_i,t^x_{i+1}\}$. Fix any solution $F$ in the instance $G^x$.
Let $A$ be the set of those indices $i$ for which $t^x_{i}b^x_{i+1}\in F$, and let $B$ be the set of those indices $i$ for which $t^x_{i+1}b^x_{i}\in F$.
Each set $W_i$ induces a $C_4$, and hence one of the edges $t^x_{i}b^x_{i+1}$ or $t^x_{i+1}b^x_{i}$ needs to be filled in $F$. 
Therefore $A\cup B=\{0,1,\ldots,p_x-1\}$.
We claim that if $i\in A$, then $i+1\notin B$. Indeed, otherwise we would obtain an induced $C_4$ with both diagonals non-fillable, which is a contradiction.
Hence, in particular $i\in A$ implies $i+1\in A$, so $A$ is either empty or equal to $\{0,1,\ldots,p_x-1\}$.
Since $i\in A$ implies $i+1\notin B$, in the latter case we have that $B$ is empty.
We conclude that either $A=\emptyset$ and $B=\{0,1,\ldots,p_x-1\}$, or $A=\{0,1,\ldots,p_x-1\}$ and $B=\emptyset$; this corresponds to the two solutions described in the statement.
\cqed\end{proof}

We now move on to the description of the clause gadget $H^c$, constructed for every clause $c\in \cls$.
Again, we use almost exactly the same construction as Drange et al.~\cite{DrangeFPV15}.
The construction is depicted in Figure~\ref{fig:c4-free-clause-gadget}, which is again taken almost verbatim from Drange et al.~\cite{DrangeFPV15}, by the consent of the authors.

\begin{figure}[t]
  \centering
  \begin{tikzpicture} [every node/.style={circle, draw, scale=.8},
    scale=.8, minimum size=1.5em, inner sep=0pt,rotate=-45]
    
    % CLAUSE!!!!
    \node (cv1) at (11,8) {$v^c_1$}; \node (cv2) at (11,6) {$v^c_2$};
    \node (cv3) at (10,7) {$v^c_3$}; \node (cv4) at (12,7) {$v^c_4$};
    
    \node (cu1) at (16,6) {$u^c_1$}; \node (cu2) at (16,8) {$u^c_2$};
    \node (cu3) at (12,2) {$u^c_3$}; \node (cu4) at (10,2) {$u^c_4$};
    
    \draw (cv1) -- (cv4) -- (cv2) -- (cv3) -- (cv1) -- (cu2) -- (cu1) -- (cv2);
    \draw (cv3) -- (cu4) -- (cu3) -- (cv4);
    
    \draw[dotted] (cv1) -- (cv2);
    \draw[dotted] (cv3) -- (cv4);
    \draw[dotted] (cv1) -- (cu1);
    \draw[dotted] (cv2) -- (cu2);
    \draw[dotted] (cv3) -- (cu3);

    \end{tikzpicture}
  \caption{The clause gadget $H^c$. The dotted lines represent fillable non-edges. The figure is taken almost verbatim from Drange et al.~\cite{DrangeFPV15}, by the consent of the authors.}
  \label{fig:c4-free-clause-gadget}
\end{figure}
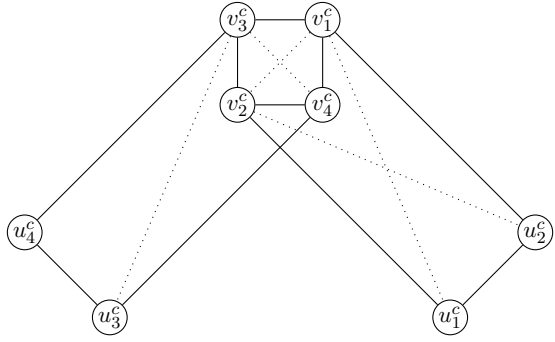

The gadget consists of $8$ vertices: $v^c_t$ and $u^c_t$, for $t=1,2,3,4$. There are five fillable non-edges: $v^c_1v^c_2$, $v^c_3v^c_4$, $u^c_1v^c_1$, $u^c_2v^c_2$, and $u^c_3v^c_3$.
All the other non-edges are declared to be not fillable; note that in particular $u^c_4v^c_4$ is not fillable. The following claim, which can be verified by a direct check, explains the properties
of the clause gadget.

\begin{claim}\label{cl:c4-compl-cls}
There is no solution in the {\sc{Sandwich $C_4$-Free Edge Completion}} $H^c$ which does not contain any of the edges $u^c_1v^c_1$, $u^c_2v^c_2$, and $u^c_3v^c_3$. However, for each $i=1,2,3$ there
is a solution $F^c_i$ that contains the edge $u^c_iv^c_i$, and does not contain any other edge from the aforementioned triple.
\end{claim}

Finally, we connect clause gadgets with variable gadgets using connector gadgets, which are just $C_4$-s. More precisely, if the $i$-th literal of a clause $c$ is $x$, and the occurrence of $x$ in
$c$ is the $(j+1)$-st occurrence of $x$ in formula $\varphi$, then we add edges:
\begin{itemize}
\item $t^x_{4j}v^c_i$ and $b^x_{4j+1}u^c_i$, provided $x$ appears in $c$ negatively; or
\item $t^x_{4j+1}v^c_i$ and $b^x_{4j}u^c_i$, provided $x$ appears in $c$ positively.
\end{itemize}
This concludes the construction of the graph $G$. The only non-edges that we allow to fill are the ones declared fillable in variable and clause gadgets: $8p_x$ diagonal non-edges in each variable gadget $G^x$,
and $5$ non-edges in each clause gadget $H^c$. All the other non-edges are declared to be non-fillable. Obviously, $G$ has $\Oh(n+m)$ vertices, edges, and fillable non-edges. 
To see that the constructed instance has the structural property asserted in the lemma statement, observe that the graph spanned by fillable non-edges consists of paths of length $2$ or $3$
and cycles of length $8$ or more, and hence it has no $C_4$ subgraph. We are left with verifying the correctness of the reduction.

First, suppose the input formula $\varphi$ has a satisfying assignment $\alpha$. Construct solution $F$ as follows:
\begin{itemize}
\item For each variable $x\in \vars$, add to $F$ the solution $F^x_{\alpha(x)}$ in the variable gadget $G^x$, given by Claim~\ref{cl:c4-compl-var}.
\item For each clause $c\in \cls$, arbitrarily choose an index $i_c\in \{1,2,3\}$ of any of its literal that satisfies it under $\alpha$; such literal exists due to $\alpha$ being a satisfying assignment.
Then add to $F$ the solution $F^c_{i_c}$ in the clause gadget $H^c$, given by Claim~\ref{cl:c4-compl-cls}.
\end{itemize}
It can be easily verified, using the fact that assignment $\alpha$ satisfies $\varphi$, that $G+F$ is $C_4$-free and hence $F$ is a solution. 
This check is also contained in Drange et al.~\cite{DrangeFPV15} (see the proof of Lemma 5.8 therein), and hence we omit it here.

For the other direction, we repeat the reasoning of Drange et al.~\cite{DrangeFPV15}, because we need to adjust it to the sandwich variant.
Suppose that there exists a subset $F$ of fillable non-edges such that $G+F$ is $C_4$-free.
By Claim~\ref{cl:c4-compl-var}, the intersection of $F$ with the fillable non-edges of each variable gadget $G^x$ has to be either equal to solution $F^x_\top$ or to solution $F^x_\bot$.
Let $\alpha\colon \vars\to \{\bot,\top\}$ be a variable assignment defined as follows: 
for a variable $x$, if the aforementioned intersection is $F^x_\top$ then we set $\alpha(x)=\top$, and otherwise, if it is $F^x_\bot$, then we set $\alpha(x)=\bot$.
To verify that $\alpha$ is a satisfying assignment, suppose, for the sake of contradiction, that some clause $c$ is not satisfied under $\alpha$. 
By Claim~\ref{cl:c4-compl-cls}, at least one of the edges $u^c_1v^c_1$, $u^c_2v^c_2$, and $u^c_3v^c_3$ must belong to $F$, say $u^c_iv^c_i$.
Let $x$ be the variable in the $i$-th literal of $c$.
Since this literal does not satisfy $c$, by the construction of $\alpha$ we infer that the two vertices in $G^x$ that are adjacent to $u^c_i$ and $v^c_i$, are connected by a filled edge of $F^x_{\alpha(x)}$.
Hence, $u^c_i$, $v^c_i$, and these two vertices form a $C_4$ in $G+F$, with both diagonals being non-fillable. This is a contradiction with $G+F$ being $C_4$-free.
\end{proof}

The hardness of {\sc{Sandwich House-Free Edge Completion}} is established by a reduction from {\sc{Sandwich $C_4$-Free Edge Completion}}.

\begin{lemma}\label{lem:qHouse-comp}
There is a polynomial-time reduction which, given an instance $G$ of {\sc{Sandwich $C_4$-Free Edge Completion}} with $n$ vertices, $m$ edges, and $k$ fillable non-edges,
with the additional assumption that the graph spanned by fillable non-edges contains no $C_4$ subgraph,
constructs an equivalent instance $G'$ of {\sc{Sandwich House-Free Edge Completion}} with $\Oh(n+m)$ vertices and edges, and $k$ fillable non-edges.
Consequently, {\sc{Sandwich House-Free Edge Completion}} is $\mathrm{NP}$-hard.
\end{lemma}
\begin{proof}
Starting from $G$, construct $G'$ as follows: for each edge $uv\in E(G)$, introduce a new vertex $w_{uv}$ and make it adjacent to $u$ and to $v$.
The fillable non-edges in graph $G'$ are only the ones that were fillable in the original instance $G$; that is, every non-edge incident to any of the new vertices is non-fillable. 
We claim that the output instance $G'$ of {\sc{Sandwich House-Free Edge Completion}} has a solution if and only if the input instance $G$ of {\sc{Sandwich $C_4$-Free Edge Completion}} has a solution.

Suppose first that $G'$ has a solution $F$. Since every non-edge incident to any vertex of $V(G')\setminus V(G)$ is non-fillable, $F$ consists only of non-edges that were fillable in the
original instance $G$. We claim that $F$ is also a solution to instance $G$ of {\sc{Sandwich $C_4$-Free Edge Completion}}. For this, it suffices to verify that $G+F$ has no induced $C_4$.
For the sake of contradiction, suppose there exists some induced $C_4$ in $G+F$, and call it $D$. 
Since in $G$ there was no $C_4$ formed by four fillable non-edges, at least one edge $uv$ of $D$ is an original edge of $G$.
For this edge we have created vertex $w_{uv}$, which is adjacent both to $u$ and to $v$. 
Since the non-edges connecting $w_{uv}$ to the other two vertices of $D$ are not fillable, we infer that $V(D)\cup \{w_{uv}\}$ induces a house in $G+F$.
This is a contradiction with $G+F$ being house-free.

For the other direction, suppose the original instance $G$ has a solution $F$; that is, $F$ consists only of fillable non-edges and $G+F$ is $C_4$-free. 
We claim that $G'+F$ is house-free, and hence $F$ is also a solution to the instance $G'$ of {\sc{Sandwich House-Free Edge Completion}}.
For the sake of contradiction, suppose $G'+F$ contains some induced house $D$; let $D'$ be the $C_4$ contained in $D$. 
At least one vertex of $D'$ does not belong to $V(G)$, because otherwise $D'$ would be an induced $C_4$ in $G+F$, which is $C_4$-free by assumption.
Hence, this vertex is of the form $w_{uv}$ for some edge $uv$ of $G$.
Note that $u$ and $v$ are the only two neighbors of $w_{uv}$ in $G'+F$, and hence they must be also its neighbors on the $4$-cycle $D'$.
However, $uv$ is an edge of $G$, which contradicts the supposition that $D'$ is an induced $C_4$.
\end{proof}

\subsection{From sandwich problems to hardness of approximation}

Having proven the NP-hardness of sandwich problems, we can use them to prove the hardness of approximation for the standard variants, as in Theorems~\ref{thm:Quar-3con} and~\ref{thm:Quar-3con-compl}.
For this, we need analogues of Theorems~\ref{thm:Quar-3con} and~\ref{thm:Quar-3con-compl}, which provide reductions from sandwich problems by turning the additional hard constraints into approximation gap.
However, the proofs of Theorems~\ref{thm:Quar-3con} and~\ref{thm:Quar-3con-compl} use the assumption about $3$-connectedness, which is not available in our current setting.
Hence, we need to verify by hand that the same strategy still works.

\begin{lemma}\label{lem:Quar-specific}
Let $(\Pi,\Pi')$ be one of the following pairs of problems: 
\begin{itemize}
\item {\sc{Sandwich $C_4$-Free Edge Deletion}} and {\sc{$C_4$-Free Edge Deletion}};
\item {\sc{Sandwich $C_5$-Free Edge Deletion}} and {\sc{$C_5$-Free Edge Deletion}};
\item {\sc{Sandwich $C_4$-Free Edge Completion}} and {\sc{$C_4$-Free Edge Completion}};
\item {\sc{Sandwich House-Free Edge Completion}} and {\sc{House-Free Edge Completion}}.
\end{itemize}
Let $p(\cdot)$ be a polynomial with $p(\ell) \geq \ell$, for all positive $\ell$. 
Then there is a polynomial time reduction, which given 
an instance $G$ of $\Pi$, creates an instance $(G',k)$ of $\Pi'$, such that: 
\begin{itemize}
 \item $k$ is the number of deletable edges, resp. fillable non-edges, in $G$;
 \item If $G$ is a YES instance of $\Pi$, then $(G',k)$ is a YES instance of $\Pi'$;
 \item If $G$ is a NO instance of $\Pi$, then $(G', p(k))$ is a NO instance $\Pi'$.
\end{itemize}
\end{lemma}
\begin{proof}
We give the proof for $(\Pi,\Pi')$ being {\sc{Sandwich $C_4$-Free Edge Deletion}} and {\sc{$C_4$-Free Edge Deletion}}, and then we shortly discuss how it can be modified to work for the other problem pairs.
Let $G$ be the input instance of {\sc{Sandwich $C_4$-Free Edge Deletion}}, and let $k$ be the number of deletable edges in $G$.
Starting from $G$, construct graph $G'$ as follows: for every undeletable edge $uv\in E(G)$, add $p(k)+2$ vertices $w_{uv}^i$, for $i=1,\ldots,p(k)+2$. 
Each of these vertices is adjacent only to $u$ and $v$. This concludes the construction of $G'$; we are left with verifying that $G'$ has the requested properties.

First, suppose that $G$ is a YES instance of {\sc{Sandwich $C_4$-Free Edge Deletion}}, that is, there is some subset $F$ of deletable edges of $G$ such that $G-F$ is $C_4$-free. 
Obviously $|F|\leq k$, because there are $k$ deletable edges in $G$ in total.
We claim that then $F$ is also a solution to instance $(G',k)$ of {\sc{$C_4$-Free Edge Deletion}}.
For this, it suffices to verify that $G'-F$ is also $C_4$-free.

For the sake of contradiction, suppose that $G'-F$ contains some induced $C_4$; call it $D$. 
Since $G-F$ is $C_4$-free, at least one vertex of $D$ is outside of $V(G)$, and hence it is of the form $w^i_{uv}$ for some undeletable edge $uv$ of $G$ and $i\in [p(k)+2]$.
As $uv$ is undeletable, we have that $uv\notin F$. As $w^i_{uv}$ has degree $2$ in $G$, we have that the two neighbors of $w^i_{uv}$ on $D$ must be $u$ and $v$.
However, $uv$ is still present in $G-F$, and hence it would be a chord in the induced $4$-cycle $D$; this is a contradiction.

For the other direction, suppose that $(G',p(k))$ is a YES instance of {\sc{$C_4$-Free Edge Deletion}}, that is, 
there is a subset $F$ of at most $p(k)$ edges of $G'$ such that $G'-F$ is $C_4$-free.

We first claim that $F$ does not contain any edge of $G$ that is undeletable.
Suppose the contrary: there is some edge $uv$ in $F$ that is an undeletable edge of $G$.
Recall that we have constructed $p(k)+2$ vertices $w^i_{uv}$ that are pairwise non-adjacent, and adjacent to $u$ and $v$.
Since $|F|\leq p(k)$, there have to be at least two of these vertices, say $w^{i}_{uv}$ and $w^{j}_{uv}$, for which $F$ does not contain any of the edges incident to $w^{i}_{uv}$ or $w^{j}_{uv}$.
Since $uv\in F$, we infer that $\{u,v,w^{i}_{uv},w^{j}_{uv}\}$ induces a $C_4$ in $G'-F$, a contradiction.

Hence, $F$ contains no undeletable edge of $G$. Consider set $F'=E(G)\cap F$: this set contains only deletable edges of $G$, and moreover $G-F'$ has to be $C_4$-free due to being an induced subgraph of $G'-F$.
We conclude that $F'$ is a solution to the original instance $G$ of {\sc{Sandwich $C_4$-Free Edge Deletion}}.

\begin{figure}
    \centering
    \subfloat[{\sc{$C_4$ Deletion}}]{
        \def\svgwidth{0.2\textwidth}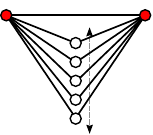
    }
    \quad
    \subfloat[{\sc{$C_5$ Deletion}}]{
	\def\svgwidth{0.2\textwidth}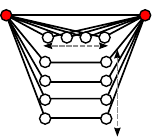
    }
    \quad
    \subfloat[{\sc{$C_4$ Completion}}]{
	\def\svgwidth{0.2\textwidth}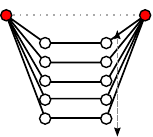
    }
    \quad
    \subfloat[{\sc{House Compl.}}]{
	\def\svgwidth{0.2\textwidth}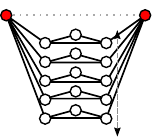
    }
    \caption{Gadgets attached to undeletable edges, resp. non-fillable non-edges, in the constructions in the proof of Theorem~\ref{lem:Quar-specific}.}\label{fig:lifting-qr}
\end{figure}

To prove the claim for the remaining $3$ pairs of problems, we need to design problem-specific gadgets that are attached to an undeletable edge, resp. non-fillable non-edge, 
to force a large cost of breaking the constraint. The constructions are given in Figure~\ref{fig:lifting-qr}. More precisely:
\begin{itemize}
\item For {\sc{Sandwich $C_5$-Free Edge Deletion}} and {\sc{$C_5$-Free Edge Deletion}}, we add $p(k)+1$ paths of length $3$ and $p(k)+1$ paths of length $2$ between $u$ and $v$, for each undeletable edge $uv$.
\item For {\sc{Sandwich $C_4$-Free Edge Completion}} and {\sc{$C_4$-Free Edge Completion}}, we add $p(k)+1$ paths of length $3$ between $u$ and $v$, for each non-fillable non-edge $uv$.
\item For {\sc{Sandwich House-Free Edge Completion}} and {\sc{House-Free Edge Completion}}, we add $p(k)+1$ paths of length $3$ between $u$ and $v$, for each non-fillable non-edge $uv$. Moreover, in each
of these paths we build a triangle on the middle edge.
\end{itemize}
It is straightforward to verify that with these constructions, essentially the same reasoning as for {\sc{$C_4$-Free Edge Deletion}} goes through. We leave the details to the reader.
\end{proof}

The only problem left is {\sc{House-Free Edge Deletion}}, which by complementation is equivalent to {\sc{$P_5$-Free Edge Completion}}. 
Note that we even did not establish hardness of the sandwich variant of this problem.
The reason for this is that we find it the simplest to prove the appropriate analogue of Lemma~\ref{lem:Quar-specific}, stated below, using a direct reduction from {\sc{Sandwich $C_4$-Free Edge Deletion}}.

\begin{lemma}\label{thm:Quar-specific-house-del}
Let $p(\cdot)$ be a polynomial with $p(\ell) \geq \ell$, for all positive $\ell$. 
Then there is a polynomial time reduction which, given 
an instance $G$ of {\sc{Sandwich $C_4$-Free Edge Deletion}} in which every $C_4$ subgraph contains an undeletable edge, creates an instance $(G',k)$ of {\sc{House-Free Edge Deletion}}, such that: 
\begin{itemize}
 \item $k$ is the number of deletable edges in $G$;
 \item If $G$ is a YES instance of {\sc{Sandwich $C_4$-Free Edge Deletion}}, then $(G',k)$ is a YES instance of {\sc{House-Free Edge Deletion}};
 \item If $G$ is a NO instance of {\sc{Sandwich $C_4$-Free Edge Deletion}}, then $(G', p(k))$ is a NO instance {\sc{House-Free Edge Deletion}}.
\end{itemize}
\end{lemma}
\begin{proof}
We perform a similar construction as in the proof of Lemma~\ref{lem:Quar-specific}. 
We start with an instance $G$ of {\sc{Sandwich $C_4$-Free Edge Deletion}}, where every $C_4$ subgraph contains an undeletable edge.
Let $k$ be the number of deletable edges in $G$.
For every undeletable edge $uv$ in $G$, we add $p(k)+2$ gadgets $Q^i_{uv}$, for $i\in [p(k)+2]$, constructed as follows.
Each gadget $Q^i_{uv}$ consists of vertices $a^{i}_{uv}$ and $b^{i}_{uv}$, and edges 
$$ua^{i}_{uv},\ ub^{i}_{uv},\ vb^{i}_{uv},\ a^{i}_{uv}b^{i}_{uv}.$$ 
The gadgets are not adjacent to each other. The construction is depicted in Figure~\ref{fig:lifting-qr-house-del}.

\begin{figure}
    \centering
        \def\svgwidth{0.4\textwidth}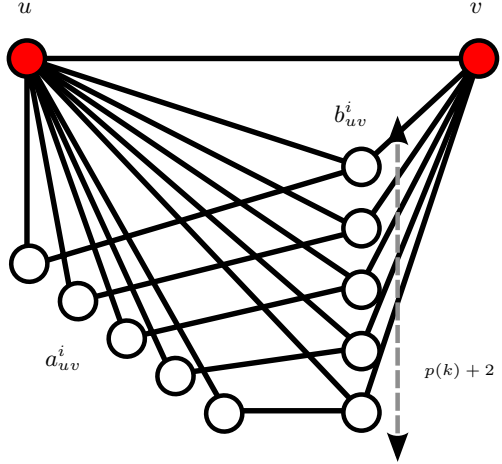
    \caption{Gadget attached to an undeletable edge in the construction in the proof of Theorem~\ref{thm:Quar-specific-house-del}.}\label{fig:lifting-qr-house-del}
\end{figure}

Let $G'$ be the obtained graph. We now verify that the construction satisfies the required properties.

Suppose first that the input instance $G$ of {\sc{Sandwich $C_4$-Free Edge Deletion}} has some solution $F$. 
That is, $F$ is a subset of deletable edges of $G$ and $G-F$ is $C_4$-free.
Obviously $|F|\leq k$, because there are $k$ deletable edges in $G$ in total.
We claim that $G'-F$ is house-free, and hence $(G',k)$ is a YES instance of {\sc{House-Free Edge Deletion}}.
For the sake of contradiction, suppose there is some induced house $D$ in $G'-F$, and let $D'$ be the induced $C_4$ contained in it.
Since $G-F$ is $C_4$-free, at least one vertex $w$ of $D'$ does not belong to $V(G)$.
Vertex $w$ cannot be of the form $a^{i}_{uv}$ for some undeletable edge $uv$, because such vertices have degree $2$ in $G-F$ and their neighbors are adjacent in $G-F$;
this cannot happen for a vertex of an induced $C_4$.
Hence, $w=b^{i}_{uv}$ for some undeletable edge $uv$ and $i\in [p(k)+2]$. 
Since $uv$ is undeletable, we have that $uv\notin F$.
We conclude that in $G-F$, the only pair of nonadjacent neighbors of $b^{i}_{uv}$ is $\{a^{i}_{uv},v\}$, and hence these must be the two neighbors of $w=b^{i}_{uv}$ on the induced $4$-cycle $D'$.
The only common neighbor of $a^{i}_{uv}$ and $v$ other than $b^{i}_{uv}$ is $u$, and hence $u$ must be the fourth vertex of the $4$-cycle $D'$.
However, $u$ and $b^{i}_{uv}$ are adjacent in $G'-F$, so $\{a^{i}_{uv},b^{i}_{uv},u,v\}$ does not induce a $C_4$ in $G'-F$. 
This is a contradiction, and we conclude that $G'-F$ is indeed house-free.

For the other direction, suppose that the instance $(G',p(k))$ of {\sc{House-Free Edge Deletion}} has some solution $F$.
That is, $F$ is a subset of edges of $G'$ of size at most $p(k)$ for which $G'-F$ is house-free.

We first claim that $F$ does not contain any edge of $G$ that was undeletable in the original instance of {\sc{Sandwich $C_4$-Free Edge Deletion}}.
Suppose the contrary: there is some edge $uv$ in $F$ that is an undeletable edge of $G$.
Recall that we have constructed $p(k)+2$ gadgets $Q^{i}_{uv}$.
Since $|F|\leq p(k)$, there have to be at least two of these gadgets, say $Q^{i}_{uv}$ and $Q^{j}_{uv}$, for which $F$ does not contain any of their edges.
Since $uv\in F$, we infer that $\{u,v,a^{i}_{uv},b^{i}_{uv},b^{j}_{uv}\}$ induces a house in $G'-F$, a contradiction.

Hence, $F$ contains no undeletable edge of $G$. 
Consider set $F'=E(G)\cap F$: this set contains only deletable edges of $G$, and we claim that it is in fact a solution to the input instance $G$ of {\sc{Sandwich $C_4$-Free Edge Deletion}}.
For the sake of contradiction, suppose $G-F'$ contains some induced $C_4$; call it $S$.
By the assumption that each $C_4$ subgraph of $G$ contains an undeletable edge, we conclude that $S$ has at least one undeletable edge, say $uv$.
Recall that for the edge $uv$ we have constructed $p(k)+2$ gadgets $Q^{i}_{uv}$.
Since $|F|\leq p(k)$, there is at least one gadget $Q^{i}_{uv}$ whose edges are disjoint with $F$.
We conclude that $S$ together with vertex $b^{i}_{uv}$ induces a house in $G'-F$, which is a contradiction.
Hence $G-F'$ is indeed $C_4$-free.
\end{proof}

Having Lemmas~\ref{lem:Quar-specific} and~\ref{thm:Quar-specific-house-del}, 
we can conclude the proof of Lemma~\ref{lem:specific} using the same reasoning as for Theorems~\ref{thm:Quar-3con} and~\ref{thm:Quar-3con-compl}.
\begin{itemize}
\item For the hardness of {\sc{$C_4$-Free Edge Deletion}}, we compose the reductions of Lemmas~\ref{lem:qC4-del} and~\ref{lem:Quar-specific} (the first problem pair).
\item For the hardness of {\sc{$C_4$-Free Edge Completion}}, we compose the reductions of Lemmas~\ref{lem:qC4-comp} and~\ref{lem:Quar-specific} (the third problem pair).
\item For the hardness of {\sc{$C_5$-Free Edge Deletion}}, we compose the reductions of Lemmas~\ref{lem:qC5-del} and~\ref{lem:Quar-specific} (the second problem pair).
The problem {\sc{$C_5$-Free Edge Completion}} is equivalent to {\sc{$C_5$-Free Edge Deletion}} by the complementation of the instance.
\item For the hardness of {\sc{$P_5$-Free Edge Deletion}}, we compose the reductions of Lemmas~\ref{lem:qHouse-comp} and~\ref{lem:Quar-specific} (the fourth problem pair)
to establish the hardness of {\sc{House-Free Edge Completion}}, and then apply the complementation of the instance.
\item For the hardness of {\sc{$P_5$-Free Edge Completion}}, we compose the reductions of Lemmas~\ref{lem:qC4-del} and~\ref{thm:Quar-specific-house-del}
to establish the hardness of {\sc{House-Free Edge Deletion}}, and then apply the complementation of the instance.
\end{itemize}
This concludes the proof of Lemma~\ref{lem:specific}, and hence of Theorem~\ref{thm:main-paths-cycles} as well.

%% file: house.pdf_tex
%% Creator: Inkscape inkscape 0.48.3.1, www.inkscape.org
%% Accompanies image file 'house.pdf' (pdf, eps, ps)
\begingroup%
  \makeatletter%
  \providecommand\color[2][]{%
    \errmessage{(Inkscape) Color is used for the text in Inkscape, but the package 'color.sty' is not loaded}%
    \renewcommand\color[2][]{}%
  }%
  \providecommand\transparent[1]{%
    \errmessage{(Inkscape) Transparency is used (non-zero) for the text in Inkscape, but the package 'transparent.sty' is not loaded}%
    \renewcommand\transparent[1]{}%
  }%
  \providecommand\rotatebox[2]{#2}%
  \ifx\svgwidth\undefined%
    \setlength{\unitlength}{103.99863281bp}%
    \ifx\svgscale\undefined%
      \relax%
    \else%
      \setlength{\unitlength}{\unitlength * \real{\svgscale}}%
    \fi%
  \else%
    \setlength{\unitlength}{\svgwidth}%
  \fi%
  \global\let\svgwidth\undefined%
  \global\let\svgscale\undefined%
  \makeatother%
  \begin{picture}(1,0.81388522)%
    \put(0,0){\includegraphics[width=\unitlength]{house.pdf}}%
  \end{picture}%
\endgroup%

%% file: c4del-variable.pdf_tex
%% Creator: Inkscape inkscape 0.48.4, www.inkscape.org
%% Accompanies image file 'c4del-variable.pdf' (pdf, eps, ps)
\begingroup%
  \makeatletter%
  \providecommand\color[2][]{%
    \errmessage{(Inkscape) Color is used for the text in Inkscape, but the package 'color.sty' is not loaded}%
    \renewcommand\color[2][]{}%
  }%
  \providecommand\transparent[1]{%
    \errmessage{(Inkscape) Transparency is used (non-zero) for the text in Inkscape, but the package 'transparent.sty' is not loaded}%
    \renewcommand\transparent[1]{}%
  }%
  \providecommand\rotatebox[2]{#2}%
  \ifx\svgwidth\undefined%
    \setlength{\unitlength}{224.8bp}%
    \ifx\svgscale\undefined%
      \relax%
    \else%
      \setlength{\unitlength}{\unitlength * \real{\svgscale}}%
    \fi%
  \else%
    \setlength{\unitlength}{\svgwidth}%
  \fi%
  \global\let\svgwidth\undefined%
  \global\let\svgscale\undefined%
  \makeatother%
  \begin{picture}(1,0.85493477)%
    \put(0,0){\includegraphics[width=\unitlength]{c4del-variable.pdf}}%
    \put(0.33828811,0.84539488){\color[rgb]{0,0,0}\makebox(0,0)[lb]{\smash{{\tn{$u_{\top}$}}}}}%
    \put(0.60519203,0.84539488){\color[rgb]{0,0,0}\makebox(0,0)[lb]{\smash{{\tn{$u_{\bot}$}}}}}%
    \put(0.40946249,0.56069737){\color[rgb]{0,0,0}\makebox(0,0)[lb]{\smash{{\tn{$v_{\top}$}}}}}%
    \put(0.55181125,0.56069737){\color[rgb]{0,0,0}\makebox(0,0)[lb]{\smash{{\tn{$v_{\bot}$}}}}}%
  \end{picture}%
\endgroup%

%% file: c4del-clause.pdf_tex
%% Creator: Inkscape inkscape 0.48.4, www.inkscape.org
%% Accompanies image file 'c4del-clause.pdf' (pdf, eps, ps)
\begingroup%
  \makeatletter%
  \providecommand\color[2][]{%
    \errmessage{(Inkscape) Color is used for the text in Inkscape, but the package 'color.sty' is not loaded}%
    \renewcommand\color[2][]{}%
  }%
  \providecommand\transparent[1]{%
    \errmessage{(Inkscape) Transparency is used (non-zero) for the text in Inkscape, but the package 'transparent.sty' is not loaded}%
    \renewcommand\transparent[1]{}%
  }%
  \providecommand\rotatebox[2]{#2}%
  \ifx\svgwidth\undefined%
    \setlength{\unitlength}{104.21059248bp}%
    \ifx\svgscale\undefined%
      \relax%
    \else%
      \setlength{\unitlength}{\unitlength * \real{\svgscale}}%
    \fi%
  \else%
    \setlength{\unitlength}{\svgwidth}%
  \fi%
  \global\let\svgwidth\undefined%
  \global\let\svgscale\undefined%
  \makeatother%
  \begin{picture}(1,0.74858287)%
    \put(0,0){\includegraphics[width=\unitlength]{c4del-clause.pdf}}%
    \put(0.2499474,0.72800371){\color[rgb]{0,0,0}\makebox(0,0)[lb]{\smash{{\tn{$s_1$}}}}}%
    \put(0.81802785,0.37487262){\color[rgb]{0,0,0}\makebox(0,0)[lb]{\smash{{\tn{$s_2$}}}}}%
    \put(0.2499474,0.006388){\color[rgb]{0,0,0}\makebox(0,0)[lb]{\smash{{\tn{$s_3$}}}}}%
    \put(0.5800482,0.72800371){\color[rgb]{0,0,0}\makebox(0,0)[lb]{\smash{{\tn{$t_1$}}}}}%
    \put(0.57237144,0.006388){\color[rgb]{0,0,0}\makebox(0,0)[lb]{\smash{{\tn{$t_2$}}}}}%
    \put(-0.00338577,0.36719586){\color[rgb]{0,0,0}\makebox(0,0)[lb]{\smash{{\tn{$t_3$}}}}}%
  \end{picture}%
\endgroup%

%% file: c4del-conn.pdf_tex
%% Creator: Inkscape inkscape 0.48.4, www.inkscape.org
%% Accompanies image file 'c4del-conn.pdf' (pdf, eps, ps)
\begingroup%
  \makeatletter%
  \providecommand\color[2][]{%
    \errmessage{(Inkscape) Color is used for the text in Inkscape, but the package 'color.sty' is not loaded}%
    \renewcommand\color[2][]{}%
  }%
  \providecommand\transparent[1]{%
    \errmessage{(Inkscape) Transparency is used (non-zero) for the text in Inkscape, but the package 'transparent.sty' is not loaded}%
    \renewcommand\transparent[1]{}%
  }%
  \providecommand\rotatebox[2]{#2}%
  \ifx\svgwidth\undefined%
    \setlength{\unitlength}{103.99863281bp}%
    \ifx\svgscale\undefined%
      \relax%
    \else%
      \setlength{\unitlength}{\unitlength * \real{\svgscale}}%
    \fi%
  \else%
    \setlength{\unitlength}{\svgwidth}%
  \fi%
  \global\let\svgwidth\undefined%
  \global\let\svgscale\undefined%
  \makeatother%
  \begin{picture}(1,0.73285623)%
    \put(0,0){\includegraphics[width=\unitlength]{c4del-conn.pdf}}%
    \put(0.26887713,0.61782739){\color[rgb]{0,0,0}\makebox(0,0)[lb]{\smash{{\tn{$u$}}}}}%
    \put(0.65349757,0.61782739){\color[rgb]{0,0,0}\makebox(0,0)[lb]{\smash{{\tn{$s$}}}}}%
    \put(0.26887713,0.04089673){\color[rgb]{0,0,0}\makebox(0,0)[lb]{\smash{{\tn{$v$}}}}}%
    \put(0.65349757,0.04089673){\color[rgb]{0,0,0}\makebox(0,0)[lb]{\smash{{\tn{$t$}}}}}%
  \end{picture}%
\endgroup%

%% file: c5del-variable.pdf_tex
%% Creator: Inkscape inkscape 0.48.4, www.inkscape.org
%% Accompanies image file 'c5del-variable.pdf' (pdf, eps, ps)
\begingroup%
  \makeatletter%
  \providecommand\color[2][]{%
    \errmessage{(Inkscape) Color is used for the text in Inkscape, but the package 'color.sty' is not loaded}%
    \renewcommand\color[2][]{}%
  }%
  \providecommand\transparent[1]{%
    \errmessage{(Inkscape) Transparency is used (non-zero) for the text in Inkscape, but the package 'transparent.sty' is not loaded}%
    \renewcommand\transparent[1]{}%
  }%
  \providecommand\rotatebox[2]{#2}%
  \ifx\svgwidth\undefined%
    \setlength{\unitlength}{232.8bp}%
    \ifx\svgscale\undefined%
      \relax%
    \else%
      \setlength{\unitlength}{\unitlength * \real{\svgscale}}%
    \fi%
  \else%
    \setlength{\unitlength}{\svgwidth}%
  \fi%
  \global\let\svgwidth\undefined%
  \global\let\svgscale\undefined%
  \makeatother%
  \begin{picture}(1,0.81099656)%
    \put(0,0){\includegraphics[width=\unitlength]{c5del-variable.pdf}}%
    \put(0.30948096,0.72627117){\color[rgb]{0,0,0}\makebox(0,0)[lb]{\smash{{\tn{$u_{\top}$}}}}}%
    \put(0.65312357,0.72627117){\color[rgb]{0,0,0}\makebox(0,0)[lb]{\smash{{\tn{$u_{\bot}$}}}}}%
    \put(0.30948096,0.50290347){\color[rgb]{0,0,0}\makebox(0,0)[lb]{\smash{{\tn{$v_{\bot}$}}}}}%
    \put(0.61875931,0.50290347){\color[rgb]{0,0,0}\makebox(0,0)[lb]{\smash{{\tn{$v_{\top}$}}}}}%
  \end{picture}%
\endgroup%

%% file: c5del-clause.pdf_tex
%% Creator: Inkscape inkscape 0.48.4, www.inkscape.org
%% Accompanies image file 'c5del-clause.pdf' (pdf, eps, ps)
\begingroup%
  \makeatletter%
  \providecommand\color[2][]{%
    \errmessage{(Inkscape) Color is used for the text in Inkscape, but the package 'color.sty' is not loaded}%
    \renewcommand\color[2][]{}%
  }%
  \providecommand\transparent[1]{%
    \errmessage{(Inkscape) Transparency is used (non-zero) for the text in Inkscape, but the package 'transparent.sty' is not loaded}%
    \renewcommand\transparent[1]{}%
  }%
  \providecommand\rotatebox[2]{#2}%
  \ifx\svgwidth\undefined%
    \setlength{\unitlength}{106.24673292bp}%
    \ifx\svgscale\undefined%
      \relax%
    \else%
      \setlength{\unitlength}{\unitlength * \real{\svgscale}}%
    \fi%
  \else%
    \setlength{\unitlength}{\svgwidth}%
  \fi%
  \global\let\svgwidth\undefined%
  \global\let\svgscale\undefined%
  \makeatother%
  \begin{picture}(1,0.81706291)%
    \put(0,0){\includegraphics[width=\unitlength]{c5del-clause.pdf}}%
    \put(0.29786484,0.79687813){\color[rgb]{0,0,0}\makebox(0,0)[lb]{\smash{{\tn{$s_1=s_2$}}}}}%
    \put(0.22256841,0.00626558){\color[rgb]{0,0,0}\makebox(0,0)[lb]{\smash{{\tn{$t_1$}}}}}%
    \put(0.59905057,0.00626558){\color[rgb]{0,0,0}\makebox(0,0)[lb]{\smash{{\tn{$t_2$}}}}}%
    \put(-0.00332089,0.4956924){\color[rgb]{0,0,0}\makebox(0,0)[lb]{\smash{{\tn{$s_3$}}}}}%
    \put(0.82493987,0.4956924){\color[rgb]{0,0,0}\makebox(0,0)[lb]{\smash{{\tn{$t_3$}}}}}%
  \end{picture}%
\endgroup%

%% file: c5del-conn.pdf_tex
%% Creator: Inkscape inkscape 0.48.4, www.inkscape.org
%% Accompanies image file 'c5del-conn.pdf' (pdf, eps, ps)
\begingroup%
  \makeatletter%
  \providecommand\color[2][]{%
    \errmessage{(Inkscape) Color is used for the text in Inkscape, but the package 'color.sty' is not loaded}%
    \renewcommand\color[2][]{}%
  }%
  \providecommand\transparent[1]{%
    \errmessage{(Inkscape) Transparency is used (non-zero) for the text in Inkscape, but the package 'transparent.sty' is not loaded}%
    \renewcommand\transparent[1]{}%
  }%
  \providecommand\rotatebox[2]{#2}%
  \ifx\svgwidth\undefined%
    \setlength{\unitlength}{103.99863281bp}%
    \ifx\svgscale\undefined%
      \relax%
    \else%
      \setlength{\unitlength}{\unitlength * \real{\svgscale}}%
    \fi%
  \else%
    \setlength{\unitlength}{\svgwidth}%
  \fi%
  \global\let\svgwidth\undefined%
  \global\let\svgscale\undefined%
  \makeatother%
  \begin{picture}(1,0.7862724)%
    \put(0,0){\includegraphics[width=\unitlength]{c5del-conn.pdf}}%
    \put(0.11502895,0.61782739){\color[rgb]{0,0,0}\makebox(0,0)[lb]{\smash{{\tn{$u$}}}}}%
    \put(0.80734575,0.61782739){\color[rgb]{0,0,0}\makebox(0,0)[lb]{\smash{{\tn{$s$}}}}}%
    \put(0.26887713,0.04089673){\color[rgb]{0,0,0}\makebox(0,0)[lb]{\smash{{\tn{$v$}}}}}%
    \put(0.65349757,0.04089673){\color[rgb]{0,0,0}\makebox(0,0)[lb]{\smash{{\tn{$t$}}}}}%
  \end{picture}%
\endgroup%

%% file: quar-c4del.pdf_tex
%% Creator: Inkscape inkscape 0.48.3.1, www.inkscape.org
%% Accompanies image file 'quar-c4del.pdf' (pdf, eps, ps)
\begingroup%
  \makeatletter%
  \providecommand\color[2][]{%
    \errmessage{(Inkscape) Color is used for the text in Inkscape, but the package 'color.sty' is not loaded}%
    \renewcommand\color[2][]{}%
  }%
  \providecommand\transparent[1]{%
    \errmessage{(Inkscape) Transparency is used (non-zero) for the text in Inkscape, but the package 'transparent.sty' is not loaded}%
    \renewcommand\transparent[1]{}%
  }%
  \providecommand\rotatebox[2]{#2}%
  \ifx\svgwidth\undefined%
    \setlength{\unitlength}{43.96810131bp}%
    \ifx\svgscale\undefined%
      \relax%
    \else%
      \setlength{\unitlength}{\unitlength * \real{\svgscale}}%
    \fi%
  \else%
    \setlength{\unitlength}{\svgwidth}%
  \fi%
  \global\let\svgwidth\undefined%
  \global\let\svgscale\undefined%
  \makeatother%
  \begin{picture}(1,0.87797602)%
    \put(0,0){\includegraphics[width=\unitlength]{quar-c4del.pdf}}%
    \put(0.02274376,0.86890138){\color[rgb]{0,0,0}\makebox(0,0)[lb]{\smash{{\tn{$u$}}}}}%
    \put(0.93249421,0.86890138){\color[rgb]{0,0,0}\makebox(0,0)[lb]{\smash{{\tn{$v$}}}}}%
    \put(0.64137407,0.15929603){\color[rgb]{0,0,0}\makebox(0,0)[lb]{\smash{{\tnn{$p(k)+2$}}}}}%
  \end{picture}%
\endgroup%

%% file: quar-c5del.pdf_tex
%% Creator: Inkscape inkscape 0.48.3.1, www.inkscape.org
%% Accompanies image file 'quar-c5del.pdf' (pdf, eps, ps)
\begingroup%
  \makeatletter%
  \providecommand\color[2][]{%
    \errmessage{(Inkscape) Color is used for the text in Inkscape, but the package 'color.sty' is not loaded}%
    \renewcommand\color[2][]{}%
  }%
  \providecommand\transparent[1]{%
    \errmessage{(Inkscape) Transparency is used (non-zero) for the text in Inkscape, but the package 'transparent.sty' is not loaded}%
    \renewcommand\transparent[1]{}%
  }%
  \providecommand\rotatebox[2]{#2}%
  \ifx\svgwidth\undefined%
    \setlength{\unitlength}{43.96810131bp}%
    \ifx\svgscale\undefined%
      \relax%
    \else%
      \setlength{\unitlength}{\unitlength * \real{\svgscale}}%
    \fi%
  \else%
    \setlength{\unitlength}{\svgwidth}%
  \fi%
  \global\let\svgwidth\undefined%
  \global\let\svgscale\undefined%
  \makeatother%
  \begin{picture}(1,0.89617103)%
    \put(0,0){\includegraphics[width=\unitlength]{quar-c5del.pdf}}%
    \put(0.02274376,0.88709639){\color[rgb]{0,0,0}\makebox(0,0)[lb]{\smash{{\tn{$u$}}}}}%
    \put(0.93249421,0.88709639){\color[rgb]{0,0,0}\makebox(0,0)[lb]{\smash{{\tn{$v$}}}}}%
    \put(0.82332416,0.15929603){\color[rgb]{0,0,0}\makebox(0,0)[lb]{\smash{{\tnn{$p(k)+1$}}}}}%
    \put(0.33205891,0.52319621){\color[rgb]{0,0,0}\makebox(0,0)[lb]{\smash{{\tnn{$p(k)+1$}}}}}%
  \end{picture}%
\endgroup%

%% file: quar-c4comp.pdf_tex
%% Creator: Inkscape inkscape 0.48.3.1, www.inkscape.org
%% Accompanies image file 'quar-c4comp.pdf' (pdf, eps, ps)
\begingroup%
  \makeatletter%
  \providecommand\color[2][]{%
    \errmessage{(Inkscape) Color is used for the text in Inkscape, but the package 'color.sty' is not loaded}%
    \renewcommand\color[2][]{}%
  }%
  \providecommand\transparent[1]{%
    \errmessage{(Inkscape) Transparency is used (non-zero) for the text in Inkscape, but the package 'transparent.sty' is not loaded}%
    \renewcommand\transparent[1]{}%
  }%
  \providecommand\rotatebox[2]{#2}%
  \ifx\svgwidth\undefined%
    \setlength{\unitlength}{43.96810131bp}%
    \ifx\svgscale\undefined%
      \relax%
    \else%
      \setlength{\unitlength}{\unitlength * \real{\svgscale}}%
    \fi%
  \else%
    \setlength{\unitlength}{\svgwidth}%
  \fi%
  \global\let\svgwidth\undefined%
  \global\let\svgscale\undefined%
  \makeatother%
  \begin{picture}(1,0.89617103)%
    \put(0,0){\includegraphics[width=\unitlength]{quar-c4comp.pdf}}%
    \put(0.02274376,0.88709639){\color[rgb]{0,0,0}\makebox(0,0)[lb]{\smash{{\tn{$u$}}}}}%
    \put(0.93249421,0.88709639){\color[rgb]{0,0,0}\makebox(0,0)[lb]{\smash{{\tn{$v$}}}}}%
    \put(0.82332416,0.15929603){\color[rgb]{0,0,0}\makebox(0,0)[lb]{\smash{{\tnn{$p(k)+1$}}}}}%
  \end{picture}%
\endgroup%

%% file: quar-house-comp.pdf_tex
%% Creator: Inkscape inkscape 0.48.3.1, www.inkscape.org
%% Accompanies image file 'quar-house-comp.pdf' (pdf, eps, ps)
\begingroup%
  \makeatletter%
  \providecommand\color[2][]{%
    \errmessage{(Inkscape) Color is used for the text in Inkscape, but the package 'color.sty' is not loaded}%
    \renewcommand\color[2][]{}%
  }%
  \providecommand\transparent[1]{%
    \errmessage{(Inkscape) Transparency is used (non-zero) for the text in Inkscape, but the package 'transparent.sty' is not loaded}%
    \renewcommand\transparent[1]{}%
  }%
  \providecommand\rotatebox[2]{#2}%
  \ifx\svgwidth\undefined%
    \setlength{\unitlength}{43.96810131bp}%
    \ifx\svgscale\undefined%
      \relax%
    \else%
      \setlength{\unitlength}{\unitlength * \real{\svgscale}}%
    \fi%
  \else%
    \setlength{\unitlength}{\svgwidth}%
  \fi%
  \global\let\svgwidth\undefined%
  \global\let\svgscale\undefined%
  \makeatother%
  \begin{picture}(1,0.89617103)%
    \put(0,0){\includegraphics[width=\unitlength]{quar-house-comp.pdf}}%
    \put(0.02274376,0.88709639){\color[rgb]{0,0,0}\makebox(0,0)[lb]{\smash{{\tn{$u$}}}}}%
    \put(0.93249421,0.88709639){\color[rgb]{0,0,0}\makebox(0,0)[lb]{\smash{{\tn{$v$}}}}}%
    \put(0.82332416,0.15929603){\color[rgb]{0,0,0}\makebox(0,0)[lb]{\smash{{\tnn{$p(k)+1$}}}}}%
  \end{picture}%
\endgroup%

%% file: quar-house-del.pdf_tex
%% Creator: Inkscape inkscape 0.48.3.1, www.inkscape.org
%% Accompanies image file 'quar-house-del.pdf' (pdf, eps, ps)
\begingroup%
  \makeatletter%
  \providecommand\color[2][]{%
    \errmessage{(Inkscape) Color is used for the text in Inkscape, but the package 'color.sty' is not loaded}%
    \renewcommand\color[2][]{}%
  }%
  \providecommand\transparent[1]{%
    \errmessage{(Inkscape) Transparency is used (non-zero) for the text in Inkscape, but the package 'transparent.sty' is not loaded}%
    \renewcommand\transparent[1]{}%
  }%
  \providecommand\rotatebox[2]{#2}%
  \ifx\svgwidth\undefined%
    \setlength{\unitlength}{43.96810131bp}%
    \ifx\svgscale\undefined%
      \relax%
    \else%
      \setlength{\unitlength}{\unitlength * \real{\svgscale}}%
    \fi%
  \else%
    \setlength{\unitlength}{\svgwidth}%
  \fi%
  \global\let\svgwidth\undefined%
  \global\let\svgscale\undefined%
  \makeatother%
  \begin{picture}(1,0.91436604)%
    \put(0,0){\includegraphics[width=\unitlength]{quar-house-del.pdf}}%
    \put(0.02274376,0.9052914){\color[rgb]{0,0,0}\makebox(0,0)[lb]{\smash{{\tn{$u$}}}}}%
    \put(0.93249421,0.9052914){\color[rgb]{0,0,0}\makebox(0,0)[lb]{\smash{{\tn{$v$}}}}}%
    \put(0.84151917,0.17749104){\color[rgb]{0,0,0}\makebox(0,0)[lb]{\smash{{\tnn{$p(k)+2$}}}}}%
    \put(0.65956908,0.68695129){\color[rgb]{0,0,0}\makebox(0,0)[lb]{\smash{{\tn{$b^i_{uv}$}}}}}%
    \put(0.07732879,0.19568605){\color[rgb]{0,0,0}\makebox(0,0)[lb]{\smash{{\tn{$a^i_{uv}$}}}}}%
  \end{picture}%
\endgroup%

%% file: conclusion.tex
\section{Conclusions}\label{sec:conc}

In this work we initiated the study of approximability of edge modification problems related to the classes of $H$-free graphs.
Mirroring known kernelization hardness results, we have shown that the problems are hard to approximate whenever $H$ is a $3$-connected graph with at least two non-edges,
or it is a long enough path or cycle. 
It therefore seems that the approximation complexity of \Hfreedelcom somewhat matches the kernelization complexity in the cases considered so far,
so it is tempting to formulate a conjecture that for every graph $H$, the \Hfreedelcom problem admits a polynomial kernel if and only if it admits a $\poly(\OPT)$-approximation algorithm.
Since neither for kernelization nor for approximability the classification is close to being complete, this conjecture should be regarded as a very distant goal.
However, one very concrete open question that arises is whether {\sc{Cograph Edge Deletion}} (equivalent to $H=P_4$) admits a $\poly(\OPT)$-approximation.
Here, we expect the answer to be positive, due to the existence of the polynomial kernel of Guillemot et al.~\cite{GuillemotHPP13}.
The same question can be asked about the diamond graph, that is, a $K_4$ minus an edge; a polynomial kernel for {\sc{Diamond-Free Edge Deletion}} was given by Cai~\cite{cai2012master}.
Also, further investigation of the links between the case of a complete graph without one edge and the \minhorn problem, seems like an interesting direction.

%% file: completion.tex
\subsection{Completion problems}
\label{sec:completion}

We first show that the complementation of a graph enables us to transfer results from the deletion setting to the completion setting. Recall that for a graph $H$,
by $\overline{H}$ we denote its {\em{complement}}, that is, a graph on the same vertex set, where two vertices are adjacent if and only if they were not adjacent in $H$.

\begin{lemma}\label{lm:complement}
Let $H$ be any graph.
Then a pair $(G,k)$ is a YES instance of \Hfreedel if and only if the pair $(\overline{G},k)$ is a YES instance of {\sc $\overline H$-free Edge Completion}.
\end{lemma}
\begin{proof} 
      The lemma follows trivially by observing that induced copies of $H$ in $G$, after complementation, are turned into induced copies of $\overline{H}$ in $\overline{G}$.
      Also, deleting edges is translated to adding edges in the complement.
\end{proof}
      Lemma~\ref{lm:complement} provides very simple reductions from \Hfreedel to {\sc $\overline H$-free Edge Completion}, and from {\sc $\overline H$-free Edge Completion} to \Hfreedel.
      Based on these, the hardness result for deletion problems from the last section can be transferred to the hardness \Hfreecom under the assumption that $\overline H$ is $3$-connected and has at least $2$ non-edges.
      This is not quite what we wanted, as Theorem~\ref{thm:main} asks for the hardness under the assumption that $H$, not $\overline{H}$, is $3$-connected and has at least $2$ non-edges. 
      For this, we employ a very similar proof strategy as before; hence, we focus on explaining the differences. We first show the hardness of the sandwich variant.
	
\begin{lemma}\label{lem:3sat-Hcom}
	Let $H$ be a $3$-connected graph with at least $2$ non-edges. 
	There is a polynomial-time reduction which, given an instance of \threesat with $n$ variables and $m$ clauses, constructs an equivalent instance $G$ of \gHfreecom 
	that has $\Oh(n+m)$ vertices, edges, and fillable non-edges. Consequently, \gHfreecom is $\mathrm{NP}$-hard for such graphs $H$.
\end{lemma}

\begin{proof}
	We use similar construction as in the proof of Lemma~\ref{thm:3sat-3conQuar}, and we change the roles of edges to non-edges. We also extend our clause gadget to cover the case with 
	$2$ non-edges in $H$, as the straightforward adaptation from the previous proof requires the existence of $3$ non-edges. 
	The gadgets are depicted in Figure~\ref{fig:H-com-gadgets}, where dotted edges are fillable, and all others are non-fillable.
	
	The variable gadget $G^x$ is obtained from $H$ by deleting any two of its edges; we label the corresponding non-edges as $e_x, e_{\neg{x}}$. 
	We forbid adding any other non-edge, thus only non-edges $e_x, e_{\nn{x}}$ can be filled within the variable gadget. Observe that filling both of them
	at the same time creates an induced copy of $H$ which cannot be destroyed, because all the other edges are non-fillable.
	
	The clause gadget $H^c$ for a clause $c = \ell_1 \vee \ell_2 \vee \ell_3 \in \cls$ is created from two copies of $H$. The first copy
	contains two labeled non-edges $e_{\ell_1}$, $e_{\ell_2 \vee \ell_3}$, corresponding to $\ell_1$, and $\ell_2 \vee \ell_3$. All other non-edges are marked as non-fillable. 
	The second copy is created from $H$ by deleting one edge, and the corresponding non-edge is identified with the non-edge $e_{\ell_2 \vee \ell_3}$ from the first copy. 
	We also pick any two other non-edges, label them as $e_{\ell_2}$ and $e_{\ell_3}$, and make all the remaining non-edges non-fillable.
	Thus, the clause gadget has only $4$ fillable non-edges: $e_{\ell_1}$, $e_{\ell_2 \vee \ell_3}$, $e_{\ell_2}$ and $e_{\ell_3}$.
	
	To see how the clause gadget works, observe that if we do not add an edge in the place of $e_{\ell_1}$, then we have to add the edge $e_{\ell_2 \vee \ell_3}$, which in turn forces us to
	fill either $e_{\ell_2}$ or $e_{\ell_3}$. This shows that at least one of three non-edges $e_{\ell_1}, e_{\ell_2}, e_{\ell_3}$ has to be filled.
	Moreover, observe that for each $i=1,2,3$, there is a solution within the clause gadget that fills only the non-edge $e_{\ell_i}$ among the aforementioned triple:
	it is either $\{e_{\ell_1}\}$ for $i=1$, or $\{e_{\ell_2 \vee \ell_3},e_{\ell_i}\}$ for $i=2,3$.
	
	The connector gadget $C$ is obtained from $H$ by (i) labeling any of its non-edges as $e_{out}$, and (ii) selecting any edge not sharing any endpoint with $e_{out}$, deleting it, and labelling the obtained
	non-edge as $e_{in}$. Such an edge not sharing any endpoint with $e_{out}$ exists due to $H$ being $3$-connected, by the same argument as we used in the proof of Lemma~\ref{thm:3sat-3conQuar}.
	We mark all other non-edges as non-fillable, thus only $e_{in}$ and $e_{out}$ can be filled. Note that filling the non-edge $e_{in}$ forces us to fill also the non-edge $e_{out}$, because
	we obtain an induced copy of $H$ that could not be destroyed otherwise.
	
      \begin{figure}
	  \centering
	  \subfloat[Variable gadget $G^x$]{
	      \raisebox{0cm}{\def\svgwidth{0.26\textwidth}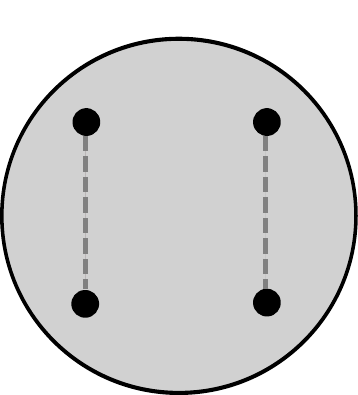}
	  }
	  \quad
	  \subfloat[Clause gadget $H^c$]{
	      \raisebox{0.5cm}{\def\svgwidth{0.3\textwidth}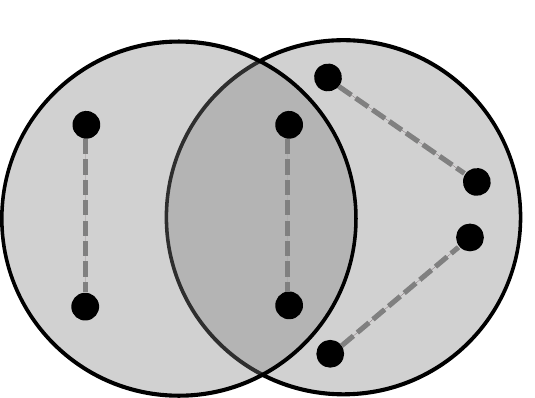}
	  }
	  \quad
	  \subfloat[Connector gadget $C$]{
	      \raisebox{0cm}{\def\svgwidth{0.23\textwidth}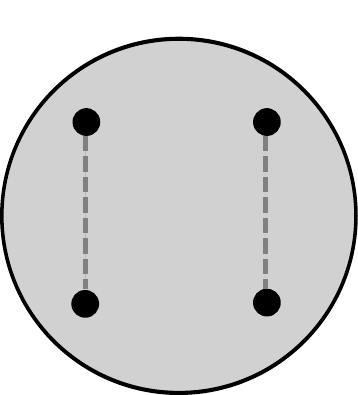}
	  }
	  \caption{Gadgets for \gHfreecom. }\label{fig:H-com-gadgets}
       \end{figure}
	
	We combine those gadgets as in Lemma~\ref{thm:3sat-3conQuar}.
	That is, the non-edges $e_{\ell_1}$, $e_{\ell_2}$, $e_{\ell_3}$ in each clause gadget $c$, are connected by chains of length $|V(H)|+2$ of connector gadgets to
	the corresponding variable gadgets. When forming the chain, the connector gadgets are attached to each other by identifying the non-edge $e_{out}$ in one gadget 
	with the non-edge $e_{in}$ in the second gadget. The chain is attached to a clause gadget by identifying the corresponding non-edge $e_{\ell_i}$ with the non-edge $e_{in}$ of the first gadget of the chain.
	Similarly, the attachment to a variable gadget is done by identifying the non-edge $e_{out}$ of the last gadget of the chain with the corresponding non-edge $e_{\ell}$ in the variable gadget.
	The explained behaviour of connector gadgets implies similar propagation of completions through the chains, as was the case for deletions in the proof of Lemma~\ref{thm:3sat-3conQuar}.
	It is easy to verify that the obtained graph $G$ has $\Oh(n+m)$ vertices, edges, and fillable non-edges, where $n$ and $m$ are the cardinalities of the variable and clause sets of $\varphi$.
	
	We have argued that the variable, clause, and connector gadgets have exactly the same functionality as in the proof of Lemma~\ref{thm:3sat-3conQuar}.
	Hence, the proof of the correctness of the reduction follows by a straightforward adaptation of the first proof; we leave checking the details to the reader.
\end{proof}

Now we show how to reduce the sandwich variant to the optimization variant by introducing a large gap. 

	\begin{lemma}\label{thm:Quar-3con-compl}
	Let $H$ be a $3$-connected graph, and $p(\cdot)$ be a polynomial with $p(\ell) \geq \ell$ for all positive $\ell$. Then there is a polynomial-time reduction which, given 
	an instance $G$ of \gHfreecom, constructs an instance $(G',k)$ of \Hfreecom such that: 
	\begin{itemize}
	 \item $k$ is the number of fillable non-edges of $G$,
	 \item 	$G'$ has $\Oh(p(k) \cdot |E(G)| \cdot |E(H)|)$ edges,
	 \item If $G$ is a YES instance, then $(G',k)$ is a YES instance,
	 \item If $G$ is a NO instance, then $(G', p(k))$ is a NO instance.
	\end{itemize}	
	\end{lemma}

	\begin{proof} 
	Similarly as in the proof of Lemma~\ref{thm:Quar-3con}, for a non-fillable non-edge $uv$, we add $p(k)$ copies of a gadget constructed as follows.
	Take $H$, arbitrarily choose one of its edges $e$, and delete $e$ from $H$. 
	The gadget is attached to the non-edge $uv$ by identifying the endpoints of $e$ with $u$ and $v$.
	The construction is presented in Figure~\ref{fig:H-del-forbid}.
	
	\begin{figure}
	  \centering
	      \raisebox{0cm}{\def\svgwidth{0.26\textwidth}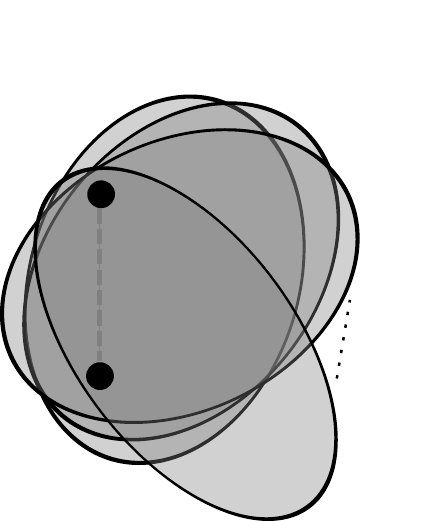}
	  
	  \caption{Gadgets $H^{uv}_i$ for \Hfreecom.}\label{fig:H-com-forbid}
       \end{figure}
	
	Observe that for any subset $F'$ of non-edges in the obtained graph $G'$, for which $G'+F'$ is $H$-free, if
	$F'$ contains the non-edge $uv$, then $F'$ also has to contain at least one non-edge within every gadget attached to $uv$.
	Otherwise the gadget would induce a copy of $H$.
	Hence, such solution $F'$ has to fill more than $p(k)$ non-edges.
	
	This shows that the functionality of the gadgets attached to non-edges is the same as in the proof of Lemma~\ref{thm:Quar-3con}.
	Consequently, a proof of correctness of the reduction follows by a straightforward adaptation of the first proof; we leave checking the details to the reader.
	\end{proof}

Exactly as in the Section~\ref{sec:deletion}, by composing the reductions of Lemmas~\ref{lem:3sat-Hcom} and~\ref{thm:Quar-3con-compl} 
we infer the hardness results promised in Theorem~\ref{thm:main} concerning completion problems.
This completes the proof of Theorem~\ref{thm:main}.

%% file: Hcom_variable.pdf_tex
%% Creator: Inkscape inkscape 0.48.3.1, www.inkscape.org
%% Accompanies image file 'Hcom_variable.pdf' (pdf, eps, ps)
\begingroup%
  \makeatletter%
  \providecommand\color[2][]{%
    \errmessage{(Inkscape) Color is used for the text in Inkscape, but the package 'color.sty' is not loaded}%
    \renewcommand\color[2][]{}%
  }%
  \providecommand\transparent[1]{%
    \errmessage{(Inkscape) Transparency is used (non-zero) for the text in Inkscape, but the package 'transparent.sty' is not loaded}%
    \renewcommand\transparent[1]{}%
  }%
  \providecommand\rotatebox[2]{#2}%
  \ifx\svgwidth\undefined%
    \setlength{\unitlength}{103.04318101bp}%
    \ifx\svgscale\undefined%
      \relax%
    \else%
      \setlength{\unitlength}{\unitlength * \real{\svgscale}}%
    \fi%
  \else%
    \setlength{\unitlength}{\svgwidth}%
  \fi%
  \global\let\svgwidth\undefined%
  \global\let\svgscale\undefined%
  \makeatother%
  \begin{picture}(1,1.10390038)%
    \put(0,0){\includegraphics[width=\unitlength]{Hcom_variable.pdf}}%
    \put(0.27679261,0.49090248){\color[rgb]{0,0,0}\makebox(0,0)[lb]{\smash{{\tn{$e_x$}}}}}%
    \put(0.58734204,0.49090246){\color[rgb]{0,0,0}\makebox(0,0)[lb]{\smash{{\tn{$e_{\neg x}$}}}}}%
    \put(0.43983121,1.03436406){\color[rgb]{0,0,0}\makebox(0,0)[lb]{\smash{{\tn{$H$}}}}}%
  \end{picture}%
\endgroup%

%% file: Hcom_clause.pdf_tex
%% Creator: Inkscape inkscape 0.48.3.1, www.inkscape.org
%% Accompanies image file 'Hcom_clause.pdf' (pdf, eps, ps)
\begingroup%
  \makeatletter%
  \providecommand\color[2][]{%
    \errmessage{(Inkscape) Color is used for the text in Inkscape, but the package 'color.sty' is not loaded}%
    \renewcommand\color[2][]{}%
  }%
  \providecommand\transparent[1]{%
    \errmessage{(Inkscape) Transparency is used (non-zero) for the text in Inkscape, but the package 'transparent.sty' is not loaded}%
    \renewcommand\transparent[1]{}%
  }%
  \providecommand\rotatebox[2]{#2}%
  \ifx\svgwidth\undefined%
    \setlength{\unitlength}{154.03009215bp}%
    \ifx\svgscale\undefined%
      \relax%
    \else%
      \setlength{\unitlength}{\unitlength * \real{\svgscale}}%
    \fi%
  \else%
    \setlength{\unitlength}{\svgwidth}%
  \fi%
  \global\let\svgwidth\undefined%
  \global\let\svgscale\undefined%
  \makeatother%
  \begin{picture}(1,0.74368178)%
    \put(0,0){\includegraphics[width=\unitlength]{Hcom_clause.pdf}}%
    \put(0.18516896,0.32840435){\color[rgb]{0,0,0}\makebox(0,0)[lb]{\smash{{\tn{$e_{l_1}$}}}}}%
    \put(0.34617646,0.32840434){\color[rgb]{0,0,0}\makebox(0,0)[lb]{\smash{{\tn{$e_{l_2 \vee l_3}$}}}}}%
    \put(0.24749443,0.69196974){\color[rgb]{0,0,0}\makebox(0,0)[lb]{\smash{{\tn{$H $}}}}}%
    \put(0.66299879,0.46344293){\color[rgb]{0,0,0}\makebox(0,0)[lb]{\smash{{\tn{$e_{l_2}$}}}}}%
    \put(0.67338769,0.22972232){\color[rgb]{0,0,0}\makebox(0,0)[lb]{\smash{{\tn{$e_{l_3}$}}}}}%
    \put(0.56950935,0.69716331){\color[rgb]{0,0,0}\makebox(0,0)[lb]{\smash{{\tn{$H \setminus e_{l_2 \vee l_3}$}}}}}%
  \end{picture}%
\endgroup%

%% file: Hcom_connector.pdf_tex
%% Creator: Inkscape inkscape 0.48.3.1, www.inkscape.org
%% Accompanies image file 'Hcom_connector.pdf' (pdf, eps, ps)
\begingroup%
  \makeatletter%
  \providecommand\color[2][]{%
    \errmessage{(Inkscape) Color is used for the text in Inkscape, but the package 'color.sty' is not loaded}%
    \renewcommand\color[2][]{}%
  }%
  \providecommand\transparent[1]{%
    \errmessage{(Inkscape) Transparency is used (non-zero) for the text in Inkscape, but the package 'transparent.sty' is not loaded}%
    \renewcommand\transparent[1]{}%
  }%
  \providecommand\rotatebox[2]{#2}%
  \ifx\svgwidth\undefined%
    \setlength{\unitlength}{103.04318101bp}%
    \ifx\svgscale\undefined%
      \relax%
    \else%
      \setlength{\unitlength}{\unitlength * \real{\svgscale}}%
    \fi%
  \else%
    \setlength{\unitlength}{\svgwidth}%
  \fi%
  \global\let\svgwidth\undefined%
  \global\let\svgscale\undefined%
  \makeatother%
  \begin{picture}(1,1.10390038)%
    \put(0,0){\includegraphics[width=\unitlength]{Hcom_connector.pdf}}%
    \put(0.27679261,0.49090248){\color[rgb]{0,0,0}\makebox(0,0)[lb]{\smash{{\tn{$e_{in}$}}}}}%
    \put(0.55628707,0.49090246){\color[rgb]{0,0,0}\makebox(0,0)[lb]{\smash{{\tn{$e_{out}$}}}}}%
    \put(0.33113876,1.03436406){\color[rgb]{0,0,0}\makebox(0,0)[lb]{\smash{{\tn{$H \setminus e_{in}$}}}}}%
  \end{picture}%
\endgroup%

%% file: Hcom_forbid.pdf_tex
%% Creator: Inkscape inkscape 0.48.3.1, www.inkscape.org
%% Accompanies image file 'Hcom_forbid.pdf' (pdf, eps, ps)
\begingroup%
  \makeatletter%
  \providecommand\color[2][]{%
    \errmessage{(Inkscape) Color is used for the text in Inkscape, but the package 'color.sty' is not loaded}%
    \renewcommand\color[2][]{}%
  }%
  \providecommand\transparent[1]{%
    \errmessage{(Inkscape) Transparency is used (non-zero) for the text in Inkscape, but the package 'transparent.sty' is not loaded}%
    \renewcommand\transparent[1]{}%
  }%
  \providecommand\rotatebox[2]{#2}%
  \ifx\svgwidth\undefined%
    \setlength{\unitlength}{124.23507717bp}%
    \ifx\svgscale\undefined%
      \relax%
    \else%
      \setlength{\unitlength}{\unitlength * \real{\svgscale}}%
    \fi%
  \else%
    \setlength{\unitlength}{\svgwidth}%
  \fi%
  \global\let\svgwidth\undefined%
  \global\let\svgscale\undefined%
  \makeatother%
  \begin{picture}(1,1.20770099)%
    \put(0,0){\includegraphics[width=\unitlength]{Hcom_forbid.pdf}}%
    \put(0.25093476,0.54472209){\color[rgb]{0,0,0}\makebox(0,0)[lb]{\smash{{\tn{$uv$}}}}}%
    \put(0.32820762,1.15002611){\color[rgb]{0,0,0}\makebox(0,0)[lb]{\smash{{\tn{$p(k) \times (H \setminus uv)$}}}}}%
    \put(0.3410866,1.01479854){\color[rgb]{0,0,0}\makebox(0,0)[lb]{\smash{{\tn{$H^{uv}_1$}}}}}%
    \put(0.56002684,1.00192004){\color[rgb]{0,0,0}\makebox(0,0)[lb]{\smash{{\tn{$H^{uv}_2$}}}}}%
    \put(0.77896598,0.83449542){\color[rgb]{0,0,0}\makebox(0,0)[lb]{\smash{{\tn{$H^{uv}_3$}}}}}%
    \put(0.79828413,0.14547921){\color[rgb]{0,0,0}\makebox(0,0)[lb]{\smash{{\tn{$H^{uv}_{p(k)}$}}}}}%
  \end{picture}%
\endgroup%